\documentclass[
  aps,
  prb,
  reprint,
  amsmath,amssymb,
  superscriptaddress
]{revtex4-2}

\usepackage{graphicx}%
\usepackage{cleveref}
\usepackage{dcolumn}%
\usepackage{bm}%
\usepackage{pgfplots}
\usepackage{filecontents}
\usepackage{geometry}
\usepackage{calc}
\usepackage{adjustbox}

\pgfplotsset{major grid style={dashed,gray}}

\newcommand{\ket}[1]{\ensuremath{| #1\rangle}}

\newcommand{\erwartung}[1]{\ensuremath{\langle #1\rangle}}

\begin{document}

\title{Time evolution of the local density of states of strongly correlated fermions coupled to a nanoprobe}
\author{T. Blum}
\affiliation{Fachbereich Physik, Philipps-Universit\"at Marburg, 
	35032 Marburg, Germany}
\author{R.M. Noack}%
\affiliation{Fachbereich Physik, Philipps-Universit\"at Marburg, 
	35032 Marburg, Germany}
\author{S.R. Manmana}%
\affiliation{Institut f\"ur Theoretische Physik, Georg-August-Universit\"at G\"ottingen, 37077 G\"ottingen, Germany}

\date{\today}
\begin{abstract}	
We study the time evolution of a one-dimensional system of strongly
correlated electrons (a `sample') that is suddenly coupled to a
smaller, initially empty 
system (a `nanoprobe'), which can subsequently move along the system.
Our purpose here is to study the role of interactions %
in this model system when it is far from equilibrium. 
We therefore take both the sample and the nanoprobe to be described by
a Hubbard model with on-site repulsive interactions and nearest-neighbor hopping.
We compute the behavior of the local particle density and the local
density of states (LDOS) %
as a function of time using time-dependent matrix product states
at
quarter and at half filling, fillings at which the chain realizes a
Luttinger liquid or a Mott insulator, respectively.
This allows us to study in detail the oscillation of the particles
between the sample and the nanoprobe. %
While, for noninteracting systems, the LDOS is time independent, in
the presence of interactions, %
the backflow of electrons to the sample will lead to nontrivial dynamics in the LDOS.
In particular,
studying the time-dependent LDOS allows us to study how the Mott gap
closes locally and how this melting of the Mott insulator propagates
through the system in time after such a local perturbation---a
behavior that we envisage can be investigated in future experiments
on %
ultrashort time scales
or on optical lattices using microscopy setups. %
\end{abstract}

\maketitle

\section{Introduction}

One very common and interesting scenario in physics occurs when two
systems are coupled to one another starting at a particular point in
time and then %
interact with one another in a time-dependent way.
The salient physical question is how the two systems then evolve in
time.
When the two systems are quantum many-body systems, particularly interesting behavior can be expected.

The study of such nonequilibrium correlated systems is inspired by
fundamental 
issues such as the nature of thermalization in closed quantum 
systems~\cite{Nature_Rigol2008,Review_Huse2015,Review_Thermalization,Review_Ueda},
by the behavior found in controlled experimental investigations in cold
atomic gases on optical lattices
\cite{RMPBloch08,review_coldgases_science}, and by the study of
material properties after a strong photo excitation, typically in the
context of pump-probe experiments
\cite{Science_Zewail2007,RMPKrausz09,PRL_Pruschke2009,Freericks_2015}.
These advanced time-resolved experimental
techniques make it possible to study  a variety of
nonequilibrium phenomena, such as the formation of
transient order, light-induced phase transitions, or hidden
states
on ultra-short time scales (pico- or femtoseconds)~\cite{PRL_Demsar2009,NaturePhysics_Ropers2018,PRL_Ravy2017,Science_Mihailovic2014,NatureComm_Mathias2016,NatureComm_Hellmann2012,Nature_Rohwer2011,Science_Schmitt2008,PRL_Singer2016,NatureComm_Rettig2016,PRL_Eggebrecht2017,Nature_Rini2007,Science_Collet2003,Science_Fausti2011}.
One such experimental approach is time- and angle-resolved photo
emission spectroscopy (trARPES)~\cite{Review_tARPES}, which provides
direct insights into nonequilibrium properties of spectral functions,
e.g., the existence of Floquet states
\cite{Floquet_firstexperiments_2013,Mahmood2016,Review_Oka_Kitamura2019,Review_properties_on_demand,RMPdelaTorre2021,Review_Fiete_2021,Subcycle_Floquet_Experiments,Floquet_Graphene_Goettingen,Floquet_Graphene_MIT},
which have been proposed as a means of
engineering certain material properties
such as topological states
in periodically driven systems \cite{Review_Eckardt2017,Review_Oka_Kitamura2019,Review_Fiete_2021}.
 
Typically, trARPES measurements are made in momentum space;
in order to gain insights into the behavior of
observables in real space, one would thus need to
Fourier transform the
data~\cite{Nature_Schmitt22}.
For locally restricted excitations, e.g., ones induced by the tip of a scanning tunneling microscope (STM)~\cite{RMP_STM,STM_collection}, it can be important to investigate the evolution of
spectral properties directly in position space.
However, STM experiments typically deal with transport properties
in the linear-response regime and hence remain close to equilibrium
(see, e.g., Ref.~\onlinecite{RMP_transport}  
for a recent review).
Here, we go beyond the equilibrium
scenario and ask what happens when the
system is perturbed so that it moves far from equilibrium.
We treat this case by computing the
time evolution of the \textit{local} density of states (LDOS) when strongly  perturbing a strongly correlated system only
locally.
We envisage that this quantity will be
experimentally
accessible in the future,
also on the ultrashort time scales treated by us
here.
Experimental realizations of such strong nonequilibrium scenarios
may be realized, for example, in experiments on optical lattices by
studying the dynamics using single-site microscopes, or in future
time-resolved local spectroscopy experiments using
STM~\cite{Review_timedependent_STM} or trARPES setups (see, e.g.,
Ref.~\onlinecite{energylandscapes_excitons},
in which
local energy landscapes in a WSe$_2$/MoS$_2$ heterostructure
are measured using trARPES techniques).

Inspired by these considerations, we ask the question of what happens when
a small, initially empty, test system (in the following referred to as
a `nanoprobe' or `probe') is brought near %
a sample hosting a strongly correlated state of matter.
We assume a strong
coupling between the sample and the nanoprobe, which also allows for
electrons to flow from the probe back to the sample. 
This will induce a dynamics that is far from the linear-response
regime, so that an equilibrium description does not apply. 
One important aspect is to what extent measures from linear-response theory are useful to describe such a situation.
We quantify this by considering in detail the properties of the LDOS computed from time-dependent single-particle propagators.
The behavior of the LDOS is particularly
interesting for strongly correlated electrons because the time-evolved
LDOS can exhibit nontrivial features induced by significant
electron-electron interaction: 
whereas the LDOS remains time independent after the sudden coupling for noninteracting systems (as discussed later in the
paper), this is no longer true for strongly interacting systems,  and the behavior of the LDOS can
change significantly in the course of time. 

Here we study these aspects on a `standard' model for strongly correlated physics, namely, the Hubbard model in one spatial dimension~\cite{Hubbard1963,Essler2005}.
At finite repulsive interaction strength, this model exhibits Mott-insulating behavior~\cite{Gebhard1997} at half filling and is a Luttinger liquid~\cite{Giamarchi2003} otherwise.
Our setup allows us to
study and compare the time evolution of the LDOS in the two
qualitatively different phases by tuning the initial
  filling.

The remainder of the paper is organized as follows: In
Sec.~\ref{sec:methods}, we describe the model, the quantities
that we study, and the
methods that we use. 
In particular, in Sec.~\ref{subsec:model}, we describe the setup of our 
model system
and the time-dependent coupling between sample and nanoprobe. 
In Sec.~\ref{subsec:observables}, we discuss the
observables we compute, in particular the time-dependent LDOS.
Section~\ref{subsec:MPSmethods}
describes our matrix-product-state (MPS) %
approach to simulating %
the dynamics of the system as well as the exact solution for the $U=0$ case.
In Sec.~\ref{sec:results} we present our results for both the stationary and the moving nanoprobe. 
Finally, we discuss and summarize our findings in Sec.~\ref{sec:summary}.
In addition, App.~\ref{subsec:convergence} describes
estimates of the accuracy of our calculations by comparing to the
exact results at $U=0$, and App.~\ref{app:v_one} contains additional results
for a higher nanoprobe velocity, $v=1$.
App.~\ref{app:u_rho} provides a theoretical estimate of the velocity
of the light-cone-like perturbation in the local electron
density in the sample that occurs after it is coupled to the probe;
this estimate is based on the exactly calculated charge velocity of
the one-dimensional Hubbard model.
We also provide supplemental material
~\cite{supplemental_material} containing additional details on the
issue of convergence of the TDVP-based time-dependent density matrix renormalization group (DMRG)
calculations for the interacting system as well as an elucidation of
the origin and form of oscillations in the LDOS for the noninteracting
system, as seen in Sec.~\ref{sec:results}.

\section{Model, Observables, and Methods}
\label{sec:methods}
\subsection{Model}
\label{subsec:model}
\begin{figure}[ht]
	\begin{tikzpicture}[scale=1] 
		\def\N{11}  
		\def\M{4} 
		\def\W{0.5}
		\def\H{1} 
		\path (-2*\W,0) -- node[auto=false]{\ldots} (0,0);
		\draw (-0.4*\W,0) -- (\N*\W+0.4*\W,0) ;
		\path (\N*\W,0) -- node[auto=false]{\ldots} (\N*\W+2*\W,0);
		\node[below=1pt] at (\N*0.5*\W-0.5*\W,0){$U$};
		\node[above=0pt] at (.5*\N*\W-0.5*\M*\W+\M*\W+2.5*\W,\H){$v$};
		\node[above=0pt] at (.5*\N*\W-0.5*\M*\W+0.5*\W,\H){$t_h$};
		\node[below=1pt] at (\N*0.5*\W-2*\W,0){$t_h$};
		\node[left=0pt] at (.25*\N*\W-0.25*\M*\W+.25*\N*\W-0.25*\M*\W-.25*\W,\H*0.5){$t'_h$};
		\foreach \x in {0,...,\N}{
			\fill[blue] (\x*\W,0) circle[radius=3pt];
		}
		\draw (.5*\N*\W-0.5*\M*\W,\H) -- (.5*\N*\W-0.5*\M*\W+\M*\W,\H) ;
		\draw [-to, line width=0.4mm](.5*\N*\W-0.5*\M*\W+\M*\W+1*\W,\H) -- (.5*\N*\W-0.5*\M*\W+\M*\W+4*\W,\H);	
		\foreach \x in {0,...,\M}{
			\draw [densely dotted] (.5*\N*\W-0.5*\M*\W+\x*\W,\H) -- (.5*\N*\W-0.5*\M*\W+\x*\W-.5*\W,0) ;
			\draw [densely dotted] (.5*\N*\W-0.5*\M*\W+\x*\W,\H) -- (.5*\N*\W-0.5*\M*\W+\x*\W+.5*\W,0) ;          
		}	
		\foreach \x in {0,...,\M}{
			\fill[red] (.5*\N*\W-0.5*\M*\W+\x*\W,\H) circle[radius=3pt];
		}	
	\end{tikzpicture}\caption{
          Schematic depiction of the setup.
          Blue sites are in the sample and red sites %
          in the nanoprobe. 
          The tunneling strength $t_h$ and the on-site Hubbard $U$ have each the same values in the sample and the probe, respectively.
          The tunneling strength $t_h'$ between sample and nanoprobe is finite only at times $t>0$.
          The nanoprobe can move to the right with a constant speed $v \geq 0$ relative to the sample.}        
        \label{fig:sketch}
\end{figure}
We study the one-dimensional Hubbard model~\cite{Hubbard1963,Essler2005}, which, at time $t=0$, is
suddenly coupled to a set of initially empty interacting sites, which represent the nanoprobe, as depicted in Fig.~\ref{fig:sketch}.
This system is modeled by the Hamiltonian 
\begin{widetext}
\begin{equation}\label{eq:hubbard-hamiltonian-tip}
	\begin{split}	
	  H=&-t_h\sum_{\ell\in\left\{\text{sample},\text{probe}\right\}}\sum_{\sigma}\left(c^\dagger_{\ell,\sigma}c_{\ell+1,\sigma}^{\phantom{\dagger}}+c^\dagger_{\ell+1,\sigma}c_{\ell,\sigma}^{\phantom{\dagger}}\right)+\,U\sum_{\ell\in\left\{\text{sample},\text{probe}\right\}}n_{\ell,\uparrow}n_{\ell,\downarrow}\\		
		&-t_h'\sum_{i \in \text{sample}}\sum_{j \in \text{probe}}\sum_{\sigma}w(t,i,j)\left(c^\dagger_{i,\sigma}c_{j,\sigma}^{\phantom{\dagger}}+c^\dagger_{j,\sigma}c_{i,\sigma}^{\phantom{\dagger}}\right)\,,
	\end{split}
\end{equation}
with
\begin{equation}
	\begin{split}
	w(t,i,j)=\Theta(t)\left[\Theta\left(vt-i+j+\frac{L-L_p}{2}+1\right)-\Theta\left(vt-i+j+\frac{L-L_p}{2}-1\right)\right]
        \, , \\
	\end{split}
\label{eq:wfunction}
\end{equation}
\end{widetext}
where $\Theta(x)$ is the Heaviside function, $t$ the time,
and we take
the lattice constant to be 1.
We designate the number of sites in the sample $L$ and the
number of sites in the nanoprobe $L_p$. Here index $i\in\{1,...,L\}$ runs over sites in the
sample and index $j\in\{1,...,L_p\}$ over sites in the probe.
We work in the canonical
ensemble with a fixed number of particles $N$.
In Hamiltonian \eqref{eq:hubbard-hamiltonian-tip}, we assume open
boundary conditions. 
The function $w(t,i,j)$ causes each site of
the nanoprobe to be coupled to the
two nearest sites in the sample only for all times $t>0$.
The operator
$c_{i,\sigma}^{(\dagger)}$ represents the annihilation (creation) operator for an electron with spin $\sigma$ on lattice site $i$, and $n_{i,\sigma} = c^\dagger_{i,\sigma}c^{\phantom{\dagger}}_{i,\sigma}$ is the electron density for spin $\sigma$ at site $i$.
Here $t_h$ is the hopping amplitude inside the sample and inside the
nanoprobe; in the following, we work in units in which $t_h \equiv 1$ and take $\hbar \equiv 1$.
The coupling $t_h'$ is the hopping amplitude between
sample and probe. 
The parameter $U \geq 0$ denotes the strength of the repulsive on-site interaction between the electrons, which we assume to be the same in the sample and in the probe. 
If not stated explicitly otherwise, we
take $U=4$,
a value corresponding to the bandwidth of the sample at $U=0$.  
In the following, we will also study the situation
in which the nanoprobe moves with a constant speed $v$ over the sample.
For speeds $v>0$, the hoppings between the nanoprobe and the sample are adjusted in time according to the function $w(t,i,j)$
which leads to a sequence of quantum quenches.

\subsection{ Observables}
\label{subsec:observables}

\subsubsection{Local particle density \erwartung{n_i}}
In all cases, we compute the time evolution of the local particle density $n_i(t) \equiv \erwartung{\psi(t)|n_i|\psi(t)}$. 
Due to the strong tunneling between sample and nanoprobe, we expect the backflow of electrons to lead to nontrivial behavior in this quantity.
Furthermore, coupling the nanoprobe to the sample is a local quench,
so that we expect that a perturbation will propagate through the sample, leading
to a light-cone-like signature in $n_i(t)$. 
Note that for the Mott insulator at half filling, the flow of electrons from the sample to the nanoprobe will move the sample away from half filling, so that it should then lose the properties of a Mott insulator in the course of time. In particular, this should affect the Mott gap, which is visible in the LDOS.

\subsubsection{Time-dependent local density of states}
We calculate the time evolution of the energy-resolved LDOS at
position $i$, $D_\sigma(t,\omega,i)$, as the Fourier transform of the retarded two-time Green's function (see, e.g., Ref.~\onlinecite{Osterkorn2022} for a similar approach to computing the nonequilibrium single-particle spectral function), 
\begin{widetext}
	\begin{equation}
		D_\sigma(t,\omega,i)=\text{Re} \left[ 
		\frac{1}{\sqrt{2\pi}}
		\int_{-\infty}^{\infty} \Theta(\tau) \, \mathrm{d}\tau\,W(\tau)e^{i\omega\tau}\left\langle\left\{ c^{\phantom{\dagger}}_{i,\sigma}(t+\tau),c_{i,\sigma}^\dagger(t)\right\}\right\rangle \right]\,,
		\label{eq:LDOS}
	\end{equation}
\end{widetext}
where the operator
$c_{i,\sigma}^{(\dagger)}(t)$ is the annihilation (creation)
operator for an electron with spin $\sigma$ at position $i$ in the
Heisenberg picture evolved with the full, time-dependent Hamiltonian
$H(t)$ at time $t$, and $\{\cdot,\cdot\}$ denotes the usual
anticommutator.
Note that alternative ways of computing momentum- and
energy-resolved time-dependent spectral functions
that avoid carrying out a
Fourier transform to frequency space do exist; see, e.g., %
Ref.~\onlinecite{Zawadzki_Feiguin}.  
Here, however, we study the local spectral properties
using the more direct definition of the LDOS based on linear-response
theory,
Eq.~\eqref{eq:LDOS},
	and compare the results to the equilibrium
expectations, giving us a quantitative measure for how strongly our time-dependent LDOS deviates from
the linear-response regime.

As discussed further below, we can compute the
two-time Green's function up to a restricted maximal
time $\tau_{\text{max}}$, so that it is necessary to regularize
the integral by introducing either a damping term
$e^{-\eta \tau}$ or a
windowing function in order to avoid ringing effects caused by a
sudden cut-off of the data at
$\tau=\tau_{\text{max}}$  \cite{numerical_recipes}. 
We find here that applying the Hann window, which is defined as
\[
 W(\tau) =
 \Theta(\tau_{\text{max}}-|\tau|)\sin^2\left(\frac{\pi\tau}{2\tau_{\text{max}}}+\frac{\pi}{2}\right)
 \,,
 \label{eq:NEqLDOS}
 \]
leads to the best results.
Note that, at equilibrium (linear-response theory), the time evolution
of $c_{i,\sigma}^{(\dagger)}(t)$ is
carried out using the Hamiltonian of the unperturbed system, so that
time-translation %
invariance can be used to reduce the time
dependence of the Green's function to one time variable.
Since time-translation %
invariance is absent out of equilibrium (e.g., in the quench we perform here at time $t=0$), we need to treat the full time dependence of the two-time Green's function.
Here
we use a relative-time representation of the retarded two-time Green's
function, i.e.,
$\left\langle\left\{c_{x,\uparrow}^{\phantom{\dagger}}(t+\tau),c_{x,\uparrow}^\dagger(t)\right\}\right\rangle$.
This has two helpful aspects: \textit{(i)} we can interpret the variable $t$ as the waiting time after switching on the coupling between the nanoprobe and the sample and ask for the behavior of the LDOS at this waiting time. 
\textit{(ii)} It makes the numerical evaluation of this quantity easier. 
This is the case because, due to the Heaviside function $\Theta(\tau)$, we only need to compute the Green's function for $\tau >0$. 
Rather than this relative-time representation, one could instead use
the so-called Wigner representation (see, e.g., Ref.~\onlinecite{Kalthoff1}), for which, however,
one would also need to compute expectation values at further points in
time, increasing the computational %
cost. 
According to Ref.~\onlinecite{Kalthoff1}, the results in both representations do not differ significantly, so that we choose to work in the relative-time representation for the sake of efficiency.

Due to the limitations of the numerical approaches, we have evaluated
the integral numerically using a
time step $d\tau = 0.01$ up to times $\tau_{\text{max}} = 5$ (except
that we take $\tau_{\text{max}} = 20$ for the $U=0$ case, which can be
treated exactly).
However, we discretize $\omega$ at 2000 equally spaced points between
$\omega_{\text{min}}=-10$ and $\omega_{\text{max}}=10$ by
interpolation using zero-padding.
Note that, at equilibrium, the LDOS defined in this way can be
expressed as a Lehmann representation with positive weights, so that an interpretation as a spectral weight is natural.
However, this is no longer possible
in the nonequilibrium case, and negative
weights can, in principle, appear~\cite{Kalthoff2}.
As discussed further below, we observe negative weights only at short
times, at which the system is strongly out of equilibrium, but, at later times, negative weights seem to be absent, so that 
an interpretation of the results in terms of spectral weights is possible.
Similarly, in nonequilibrium situations, one needs to be careful about identifying the occupied and unoccupied parts of the LDOS: at equilibrium, one usually introduces the lesser and the greater parts of the LDOS, $B_\sigma^<(\omega,i)$ and $B_\sigma^>(\omega,i)$ \cite{Meyer2022}, which indicate the populated and empty states on lattice site $i$, respectively. 
Out of equilibrium, one can
generalize these functions as follows:
\begin{widetext}
	\begin{equation}
	  B_{\sigma}^<(t,\omega,i)=\text{Re}\,\left [
            \frac{1}{\sqrt{2\pi}}\int_{0}^{\tau_{\text{max}}}\mathrm{d}\tau\,W(\tau)e^{i\omega\tau}\left\langle
                  c_{i,\sigma}^\dagger(t)
                  c^{\phantom{\dagger}}_{i,\sigma}(t+\tau)\right\rangle
                  \right ] \,,	
		\label{eq:Bsmaller}
	\end{equation}
	
	\begin{equation}
		B_{\sigma}^>(t,\omega,i)=\text{Re}\,\left
                [\frac{1}{\sqrt{2\pi}}\int_{0}^{\tau_{\text{max}}}\mathrm{d}\tau\,W(\tau)e^{i\omega\tau}\left\langle
                  c^{\phantom{\dagger}}_{i,\sigma}(t+\tau)c_{i,\sigma}^\dagger(t)\right\rangle
                  \right ] \,.
		\label{eq:Bgreater}
	\end{equation}
\end{widetext}
Note that, due to the lack of a Lehmann representation, both quantities can become negative. Furthermore, linear-response theory requires time-translation invariance in order to interpret these quantities as occupied or empty parts of the spectrum, respectively. Here, due to the nonequilibrium setup, we do not have this symmetry when going to negative $\tau$, so that we can only expect these quantities to represent the occupied (empty) part of the LDOS approximately.
We will come back to this interesting aspect in Sec.~\ref{subsec:discrepancies}, where we discuss how to estimate the occupied states in $D^<_{\sigma}(t,\omega,i)$ at lattice site $i$ by comparing to $\langle n_{i,\sigma}\rangle(t)$.
\subsubsection{Nonequilibrium occupation energy}
At equilibrium, the LDOS is obtained as
\begin{eqnarray}
	D^{\rm eq}_\sigma(\omega,i) &=& \text{Re} \left[ 
	\frac{1}{\sqrt{2\pi}}
	\int_{-\infty}^{\infty} \Theta(\tau) \, \mathrm{d}\tau\,W(\tau)e^{i\omega\tau} \right. \nonumber \\ 
	&&\left.\left\langle\left\{ c^{\phantom{\dagger}}_{i,\sigma}(\tau)^{H_0},c_{i,\sigma}^\dagger(0)^{H_0}\right\}\right\rangle \right]\,,
	\label{eq:LDOS_eq}
\end{eqnarray}
where $H_0$ indicates that the time evolution of the operators is
carried out using the unperturbed Hamiltonian, and, due to
time-translation %
invariance, only one time variable needs to be treated.
Integrating the equilibrium LDOS $D^{\rm eq}_\sigma(\omega,i)$ in energy up to the Fermi level $E_F$ will 
yield the local particle density
$\langle n_{i,\sigma}\rangle$, so that the LDOS below  $E_F$ can be interpreted as the occupied part.
The Fermi energy $E_F$ is hence the value of the energy of the highest populated state in the system.
Similarly, at equilibrium, one can introduce
\begin{eqnarray}
	B^{<,\rm eq}_\sigma(\omega,i) &=& \text{Re} \left[ 
	\frac{1}{\sqrt{2\pi}}
	\int_{-\infty}^{\infty}  \, \mathrm{d}\tau\,W(\tau)e^{i\omega\tau} \right. \nonumber \\ 
	&&\left.\left\langle c^{\dagger}_{i,\sigma}(0)^{H_0} c_{i,\sigma}^{\phantom{\dagger}}(\tau)^{H_0}\right\rangle \right]
	\label{eq:Blesser_eq}
\end{eqnarray}
as the lesser part of the LDOS.
Note that $B_\sigma^{<,\rm eq}(\omega,i)$ does not have weights for energies higher than $E_F$, so that $ \int_{-\infty}^{\infty} d\omega \, B_\sigma^{<, \rm eq}(\omega,i) = \langle n_{i,\sigma}\rangle$. Hence, $B_\sigma^{<,\rm eq}(\omega,i) $ represents the occupied part of the LDOS.
Out of equilibrium, however, one needs to be more careful.
A simple way to identify the populated states is to assume that only the lowest-energy parts of the nonequilibrium LDOS $D_\sigma(t,\omega,i)$ are occupied.
Integrating $D_\sigma(t,\omega,i)$ at site $i$ at fixed waiting time $t$ up to a certain energy, which we call $E_{\rm occ}$, should give the same result as $\langle n_{i,\sigma}\rangle(t)$.  
The value $E_{\rm occ}(t)$ can then be interpreted as the nonequilibrium 
generalization of $E_F$.
We define $E_{\rm occ}$ implicitly via
\begin{equation}
	\frac{1}{C}\int_{-\infty}^{E_{\rm occ}(t,i)}\mathrm{d}\omega\,D_\sigma(t,\omega,i)\,=\erwartung{n_{i,\sigma}}(t)\,,
	\label{eq:E_occ}
\end{equation}
with the normalization $C=\int_{-\infty}^{\infty}\mathrm{d}\omega\,D_\sigma(t,\omega,i)$.
In the following, spin-flip symmetry is present; thus, for simplicity, we only discuss the $\sigma =\, \uparrow$
component of both the LDOS and the particle density.

Furthermore,
we can compare the expectation value for the number of particles as obtained from the 
energy integral $\int d\omega \, B^<_{\sigma}(t,\omega,i)$ at fixed
$t$ and $i$ with the directly computed expectation value $\langle
n_{\sigma,i}\rangle(t)$, which, in equilibrium, is identical, as
discussed above. Differences from this expectation value can be considered
to be a measure of  %
how far away from equilibrium the system is.
As discussed further below, significant deviations are obtained
at short times in particular, indicating strong nonequilibrium
behavior, while, at later times, both quantities agree to within a few percent.
The time evolution of $E_{\rm occ}(t)$ also indicates how particles are redistributed in the course of time, in particular, for the
cases in which the LDOS is changing in time due to the strong interactions
in the system.
\subsection{ MPS-based methods}
\label{subsec:MPSmethods}
\subsubsection{Initial state preparation and time evolution}
All calculations for the interacting system have been performed at $U=4$ using a DMRG algorithm~\cite{white1992,white1993} within the MPS framework~\cite{McCulloch2007,Schollwoeck2011}, which is  available in the SymMps toolkit \cite{SymMps}. 

The general procedure is as follows. 
For $U = 4$, we calculate the ground state
on a system with $L=50$ sample lattice sites and $L_p=5$
nanoprobe sites
for the quarter- and half-filled samples, 
with $N=26$ and $50$ particles, respectively.
(We take $N=26$ rather than $N=25$ particles for quarter filling so
that the number of spin-up and spin-down particles is equal.)
Initially, the nanoprobe is not yet coupled to the system, i.e., $t_h'=0$. 
To ensure a state with zero occupation at time $t=0$ in the nanoprobe, we add a small repulsive electrostatic potential $\sum\limits_{j \in \,{\rm probe}} \mu_j (n_{j,\uparrow}+n_{j,\downarrow})$ on these sites.
We then perform a ground-state search with a maximum bond dimension
of $\chi=2500$ and a maximum discarded weight of $\delta=10^{-14}$
with a total of 50 sweeps.
This allows us to approximate the ground state of the system with an
absolute error in the ground-state energy of $3.0\times10^{-6}$ in the
quarter-filled case, compared %
to the exact value known from the Bethe ansatz~\cite{Lieb1968,Essler2005}.
The absolute error in %
the ground-state energy in the half-filled case
is even smaller, $1.7\times10^{-7}$. 

At time $t=0$, the electrostatic potential in
the nanoprobe is then set to zero, and it is immediately coupled to the system with the full hopping strength, $t_h'=t_h$.
We compute the subsequent time evolution of the total system using the time-dependent variational principle (TDVP) in its two-site version using MPS \cite{Haegeman2016,Paeckel2019}.
This allows us to treat systems with arbitrary coupling ranges and to
adjust the bond dimension in the course of the time evolution, keeping track of the growth of entanglement with time. 
It is known that the TDVP can have substantial problems with product
initial states. Alternatively, one could use the matrix product
operator (MPO) WII time-evolution algorithm
\cite{Zaletel2015,Paeckel2019}, which does not suffer from
such problems.
Here we have tested both algorithms, comparing with exact
diagonalization for systems with three particles, and we find that the
errors of the TDVP for the given setup are an order of magnitude
smaller than those of the MPO WII algorithm. The MPO WII has %
a maximum absolute error of $\approx4\times10^{-4}$ in $n_i(t)$ for
this test case, whereas the TDVP algorithm reaches %
a maximum error of $\approx5\times10^{-5}$.
In addition, in App.~\ref{subsec:convergence}, we compare the time
evolution of the half-filled case at $U=0$ as obtained from MPS
with that of %
the exact solution
(see Sec.~\ref{sec:exact}). As shown there, the absolute errors grow in
time. For $v=0$, we find that the absolute errors in $\langle
n_{i,\sigma}\rangle(t)$ are $\lesssim 0.05$ at later times, and the
absolute errors in the LDOS are $\lesssim 0.015$, while for the moving probe
the errors in the LDOS are even smaller, although the absolute errors in
$\langle n_{i,\sigma}\rangle(t)$ can be as high as $\sim 0.1$
at isolated points.
Hence, we believe that our results are sufficiently accurate to extract the essential physical behavior.

We set the threshold for the discarded weight during the time evolution to $\delta=10^{-10}$ and the maximum bond dimension to $\chi_{\rm max}=2500$. 
The time step size is $\delta t=0.01$ for all results presented. 
As discussed in App.~\ref{subsec:convergence}, the resulting discarded
weight at later times is rather large, almost $10^{-4}$; however, due
to the high computational cost, we
do not further increase
the bond dimension or
carry out calculations
to longer times. 

We perform the time evolution so that, for all waiting times, the integration time is $\tau_{\text{max}} = 5$ after the waiting time (i.e., for waiting time $t=5$, we perform the simulations up to times $t+\tau_{\text{max}} = 10$).
In this way, all the presented results at different waiting times have the same resolution in $\omega$.  

For a finite speed $v>0$ of the probe, the Hamiltonian is explicitly time dependent.
In principle, more accurate results can be obtained by applying more
elaborate discretization schemes on the time variable, such as
commutator-free exponential time propagators
\cite{Alvermann2011}. 
However, in our case, due to the structure of the function $w(t,i,j)$ in the Hamiltonian~\eqref{eq:hubbard-hamiltonian-tip}, the time dependence of $H$ is not continuous, but comes in steps at points in time determined by the value of $v$.
The time evolution for $v>0$ is, therefore, realized as a sequence of quantum quenches, for which the usual methods (here the TDVP) perform well.

\subsubsection{Swap operator}
For the case of the moving nanoprobe, i.e., $v>0$, the hopping terms
to the sample become longer range with time, which leads to larger
entanglement between the two parts of the system and hence makes the
MPS description %
less accurate.
In our case, we have only nearest-neighbor and
next-nearest-neighbor hopping at time $t=0$; it is useful to keep the hopping as short-range as possible in the course of the time evolution.           
We address this by introducing the swap operator~(see Ref.~\onlinecite{Stoudenmire_2010} for the usage of swap operators for time-evolution algorithms),
	\begin{equation}\label{eq:swap-operator}	
			P_i= \prod_{\sigma = \uparrow,\downarrow} \left[ 1 - \left(c_{i,\sigma}^\dagger-c_{i+1,\sigma}^\dagger \right) \left(c^{\phantom{\dagger}}_{i,\sigma} - c_{i+1,\sigma}^{\phantom{\dagger}} \right )\right] \, ,
	\end{equation}
which is applied at the points in time when the probe, according to the function $w(t,i,j)$, Eq.~\eqref{eq:wfunction}, couples to new sites in the sample. 
In this way, the site labeling is adapted so that the hopping terms for all times treated are either nearest-neighbor or next-nearest-neighbor terms, thus minimizing the entanglement growth in the course of time. 

\subsubsection{Solution for $U=0$}
\label{sec:exact}
In the noninteracting case, $U=0$, we can solve the system exactly by
diagonalizing the Hubbard Hamiltonian
(\ref{eq:hubbard-hamiltonian-tip}) for a single particle.
In doing this, we add 
an additional chemical
potential of $\mu=10^8$ to the sites of the nanoprobe so that the
wave-function amplitude is vanishingly small there.
This ensures that the initial many-body state has zero particle
density on the nanoprobe
before the coupling between nanoprobe and sample is turned on at $t=0$.
This then yields the single-particle wave functions $\phi_{i,n}(0)$,
where $i$ denotes the lattice site and $n$ refers to the $n$-th
excited state for a single particle, with $n=1$ being the
single-particle ground state.
The time-dependent state $\ket{\psi_i(t)}$ is then easily obtained by applying the time-evolution operator $e^{-iHt}$:
	\begin{align*}
		\ket{\psi_i(t)}&=\\
		&\left(\sum_{n=1}^{N_\uparrow}\phi_{i,n}(t)c_{n,\uparrow}^\dagger + \sum_{n=1}^{N_\downarrow} \phi_{i,n}(t) c_{n,\downarrow}^\dagger \right)\ket{0},
	\end{align*}
	where $\phi_{i,n}(t)=e^{-iHt}\phi_{i,n}(0)$ are the time-evolved single-particle states. 
	The time-dependent LDOS $D_\sigma(t,\omega,i)$ can then be calculated via the transformation $c_{i,\uparrow}(t)=\sum_{n'}\phi_{i,n'}(t) \, c_{n',\uparrow}\,.$
	Note that $n$ refers to the single-particle states for the decoupled system; $n'$, however, refers to the single-particle eigenstates after coupling the nanoprobe to the Hubbard chain.
	Inserting this expression into Eq.~\eqref{eq:LDOS} leads to a
        simple expression for the two-time Green's function: %
	\begin{align}
			\left\langle\left\{c^{\phantom{\dagger}}_{i,\sigma}(t+\tau),c_{i,\sigma}^\dagger(t)\right\}\right\rangle&= \nonumber \\
			&\sum_{n'}\phi^{\phantom{\ast}}_{i,n'}(t+\tau)\phi_{i,n'}^\ast(t) \, .
		\label{eq:LDOS_explicit}
	\end{align}
This expression is independent of the number of particles in the
system, as particles and holes lead to the same contribution for the
LDOS in the $U=0$ case. 
Furthermore, at $v=0$, i.e., for a time independent Hamiltonian, the
LDOS of the noninteracting system is time-independent because the
dependence on $t$ in %
Eq.~\eqref{eq:LDOS_explicit} vanishes for
$e^{-iH(t+\tau)}\,e^{iHt}=e^{-iH\tau}$.
Note that Eq.~\eqref{eq:LDOS_explicit} can be calculated
accurately to arbitrarily long times, so that, in contrast to the MPS results
at finite $U$, we can obtain much better resolution in $\omega$.

\section{Results}
\label{sec:results}

In this section, we present representative
results for the time evolution of our setup.
We organize the discussion by first considering a stationary
nanoprobe, i.e., one that does not move
over the sample, $v=0$, then going on to a
relatively slow nanoprobe with $v=0.55$.
We have also carried out calculations at a higher velocity, $v=1$,
which are presented in App.~\ref{app:v_one}; we will also describe
the salient aspects of the behavior relative to that of $v=0.55$ in
the main text.
For all values of $v$, we first discuss the behavior of the
noninteracting case, $U=0$, then present results for the
interacting case,
taking the Hubbard interaction to have intermediate strength,
$U = 4$. %
For all cases, we take a system with $L=50$ sample sites with
$L_p=5$ initially empty probe sites and 
treat two values of the band filling, quarter filling
with $N=26$ particles
and half filling
with $N=50$ particles.
We present the results for the time evolution of the local
particle density
in the sample and in the nanoprobe as well as the LDOS at three fixed
positions: in the sample far away from the nanoprobe,
site $i_s=13$;
in the sample directly under
the nanoprobe, site $i_s=26$;
and in the center of the nanoprobe, site $i_p=3$.
All results are obtained for samples with $L=50$ lattice sites and
nanoprobes with $L_p=5$ lattice sites.
\subsection{Resting nanoprobe: $v=0$}

We first consider the case in which the nanoprobe does not move, i.e.,
$v=0$.
As soon as the tunneling between the system and the nanoprobe is turned on at time $t=0$, the system evolves nontrivially in time as it
undergoes a quantum quench.
As the tunneling will also be turned on at $t=0$ for the cases of a
moving nanoprobe, the $v=0$ behavior will provide a basis for the
interpretation of the moving cases.
Since the nanoprobe initially holds no particles, 
particles will begin tunneling from the system to the 
nanoprobe at $t=0$.

\subsubsection{Noninteracting case $U=0$}
\begin{figure}[!]	
	\includegraphics[width=\columnwidth]{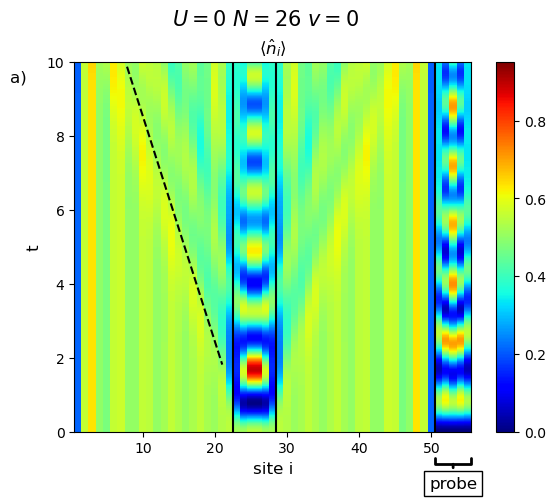}	
	\includegraphics[width=\columnwidth]{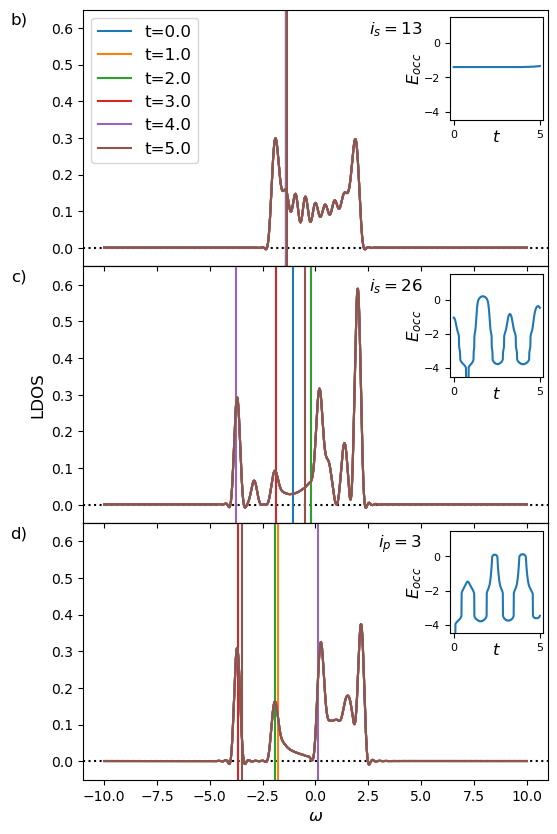}								
	\caption{
          Exact results for the nanoprobe model
          with $L=50$ sample sites and $L_p=5$ nanoprobe sites
          for the
          system initially at quarter filling, i.e., $N=26$, with
          $U=0$ %
          and nanoprobe %
          velocity $v=0$.
          Here
          (a) depicts the expectation value of the local particle
          density, $\langle {n}_i \rangle$, as a color density
          plot as a function of lattice site $i$ and time $t$,
          and 
	  (b), (c), and (d) plot the LDOS %
          for the lattice sites $i_s=13$, $i_s=26$, and $i_p=3$,
          respectively,
          as a function of the frequency $\omega$.
          Note that the LDOS %
          is rigorously time independent for $U=0$
          when $v=0$.
          The occupation energy
          $E_{\rm occ}(t)$ is shown as an inset %
          for each of the
          three lattice sites as a function of $t$ and is also
          shown at five indicated times
          as vertical lines in the LDOS.
          }
\label{fig:u0n26v0}						
\end{figure}

We start with the noninteracting case, $U=0$. 
In Fig.~\ref{fig:u0n26v0}(a), we display the expectation value of the
local particle density
$n_i(t)$ %
for the quarter-filled system
as a color density plot plotted as a function of position and time.
As can be seen in the plot, %
at time $t=0$, the nanoprobe is completely empty of particles.
After the tunneling to the system is turned on ($t>0$),
the particles begin to tunnel back and forth, %
 leading to oscillations in time in the local
particle density in both the nanoprobe and in the region of the system in its vicinity.
These oscillations are somewhat spatially inhomogeneous both in the
system and in the nanoprobe, with the strongest fluctuations of particle density
occurring near the center of the nanoprobe.
As time progresses, %
the time dependence in local
particle density spreads out in time.
In particular, %
a ``light cone'' that spreads out across the
sample at constant speed
$c\approx 1.6$, which we have estimated by 
roughly fitting the wave front, is evident.
See App.~\ref{app:u_rho} for additional discussion of the
fitting as well as a theoretical estimate of the light-cone velocity
based on the velocity of charge excitations $u_\rho$ in the system.
In Figs.~\ref{fig:u0n26v0}(b)--(d), %
we display %
the LDOS $D_\uparrow(t,\omega,x)$ for this case at
sites $i_s=13$ and $26$ in the sample and $i_p=3$ in the nanoprobe.
Sites $i_s=26$ and $i_p=3$
are directly next to each other and
become directly linked by hopping
terms as soon as the coupling between probe and sample is turned on.
For $U=0$ and a speed of $v=0$, the LDOS is time independent for all
sites after the coupling is turned on.
This is to be expected because the Hamiltonian
is time independent for $t>0$
in this case, and there is no scattering 
between the electrons; see also the discussion in
Sec.~\ref{sec:exact}. 
Thus, we display the LDOS %
for time $t=0$ only.
For $i_s=13$, Fig.~\ref{fig:u0n26v0}(b), we observe %
peaks at $\omega\approx-2$ and $2$,
with an oscillating and non-vanishing LDOS in between
the peaks.
A discussion of the origin and form of the oscillations in
the LDOS can be found in the supplemental
material~\cite{supplemental_material}.

At
sample site $i_s=26$ and probe site $i_p=3$, which are adjacent to one
another, we can see that the LDOS for both sites have peaks that are largely
at the same $\omega$ positions,  
with only the amplitudes differing somewhat.
In contrast to the LDOS
for $i_s=13$, there is a major peak at 
$\omega\approx-4$ for these two sites.
Regarding the occupation energies, we find that $E_{\rm occ}(t)$ is
essentially time independent on lattice site $i_s=13$, indicating no
(or only very little) redistribution of particles on the time
scale considered.
This is to be expected, as the light-cone-like
perturbation in $n_i(t)$ reaches this position only at later times, $t>5$, which are not shown. 
At site $i_s=26$, $E_{\rm occ}(t)$ is initially close to the equilibrium Fermi energy at site $i_s=13$. 
After the nanoprobe is coupled to the
sample, $E_{\rm occ}(t)$ begins to oscillate,
almost reaching the left edge of the 
LDOS (i.e., the lattice site is nearly completely
depleted) at time $t=4$.
The occupation energy at site $i_p=3$ inside the nanoprobe is not defined at time $t=0$ because
the system is empty. 
However, after the probe is coupled to the sample, $E_{\rm occ}(t)$ begins to oscillate, 
similarly to the value on site $i_s=26$, which is directly coupled to
this lattice site of the nanoprobe. The oscillations of $E_{\rm
  occ}(t)$ on these two sites have
approximately opposite phase, indicating strong tunneling of particles between the two lattice sites. This is to be expected, since at $U=0$ there is no scattering between the particles, which could lead to equilibration of local observables on short time scales.

\subsubsection{Interacting case at quarter filling}
\begin{figure}[!]
	\includegraphics[width=\columnwidth]{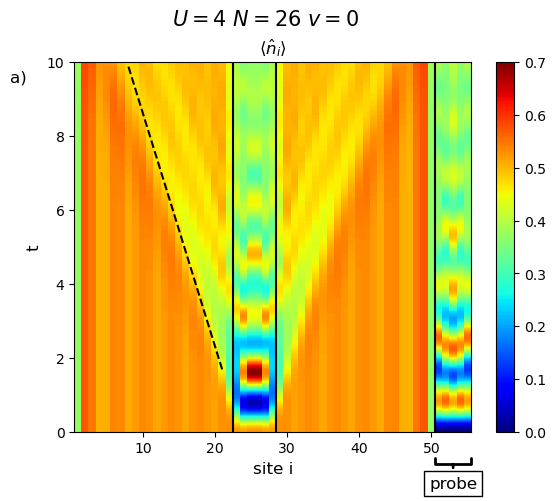}
	\includegraphics[width=\columnwidth]{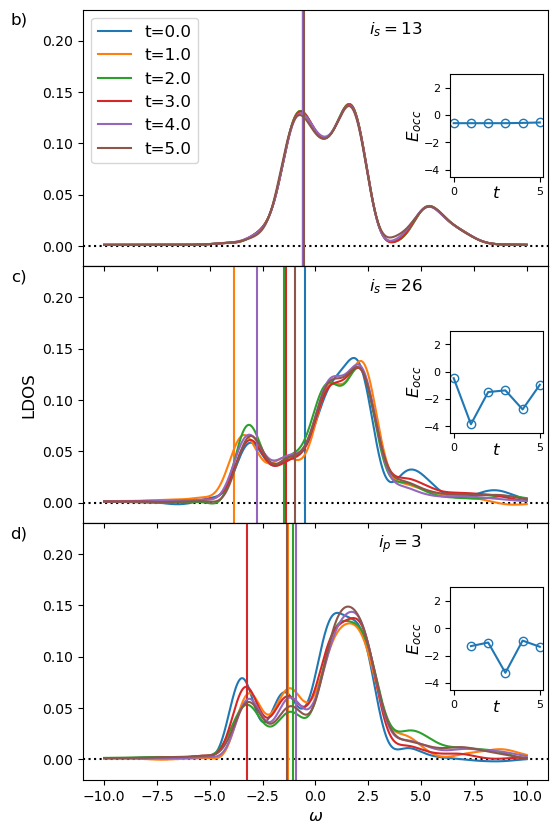}						
	\caption{
		MPS results for (a) the local particle density and the LDOS for (b) $i_s=13$, (c) $i_s=26$,
and (d) $i_p=3$, plotted as in Fig.~\ref{fig:u0n26v0}, for $U=4$,
quarter filling, i.e., $N=26$, and $v=0$.
		}\label{fig:u4n26v0}							
\end{figure}

We now turn on %
the Hubbard interaction to %
$U = 4$, %
staying  with quarter filling and
resting nanoprobe, $v=0$; see Fig.~\ref{fig:u4n26v0}.
As can be seen by comparing Fig.~\ref{fig:u4n26v0}(a) with
Fig.~\ref{fig:u0n26v0}(a), the oscillation in local particle
density
between the system and the nanoprobe is
more damped than in the $U=0$ case.
Again, %
a light cone that spreads %
out over %
the system with the
speed %
$c \approx1.6$ is clearly evident.
This value of $c$ is consistent with the theoretical estimate
in App.~\ref{app:u_rho}.
Examining the LDOS for site $i_s=13$ (well to the left of the nanoprobe), 
we obtain a double-peak structure, and an additional separated satellite peak at higher $\omega$. 
Again, as in Fig.~\ref{fig:u0n26v0}(b), $E_{\rm occ}(t)$ is essentially time independent on this lattice site on the time scales shown.
As before, this is to be expected from the light-cone signal in $n_i(t)$, which reaches this position only after the times shown. 

For lattice sites $i_s=26$, Fig.~\ref{fig:u0n26v0}(c), and $i_p=3$,
Fig.~\ref{fig:u0n26v0}(d),
the overall structure of the LDOS is similar at all times.
However, in contrast to the noninteracting case,
the LDOS is time dependent at both sites.
This is due to the interactions between the electrons.
Interestingly, at the later times treated by us, they seem to settle
to a stationary value, and the change in time is smaller
than at the beginning of the evolution. 
Again, $E_{\rm occ}(t)$ changes significantly in time on both sites and oscillates with the period of the oscillations of the local densities on these sites.
The phase of the oscillation again shows a tendency to be opposite on both sites, but the electron-electron interaction now induces a damping.

Note that, as discussed in Sec.~\ref{sec:methods}, the resolution in
$\omega$ is significantly lower than for the $U=0$ case, so that
finer structures are not resolvable.
In addition, on general grounds, the interaction $U$ leads to self-energy contributions, which tend to broaden
out peaks in spectral functions relative to the noninteracting case.

\subsubsection{Interacting case at half filling}
\begin{figure}[!]
	\includegraphics[width=\columnwidth]{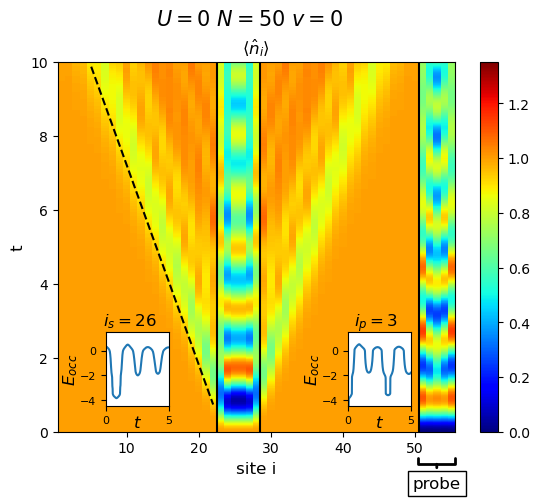}							
	\caption{Exact results for the local particle density plotted
          as in
          Fig.~\ref{fig:u0n26v0} for $U=0$, half filling, i.e., $N=50$, and $v=0$.
          The occupation energy
	  $E_{\rm occ}(t)$ is shown as an inset for sites $i_s=26$ and
          $i_p=3$.
          The LDOS in this case is not explicitly shown, as it is the
          same as in Fig.~\ref{fig:u0n26v0}.
        }\label{fig:u0n50v0}							
\end{figure}

We now consider the effect of changing the initial filling of the system to
half filling. %
Figure~\ref{fig:u0n50v0} depicts the time evolution of the local
particle density %
for the noninteracting ($U=0$) half-filled system.
In contrast to the quarter-filled case, %
Fig.~\ref{fig:u0n26v0}(a), we can see that the
oscillatory behavior is now spread
across the whole nanoprobe rather than being concentrated
in its center.
As discussed in Sec.~\ref{sec:exact}, the LDOS for the $U=0$ case is independent of band filling. However, $E_{\rm occ}$ changes in time, as shown in Fig.~\ref{fig:u0n26v0}. 

\begin{figure}[!]
	\includegraphics[width=\columnwidth]{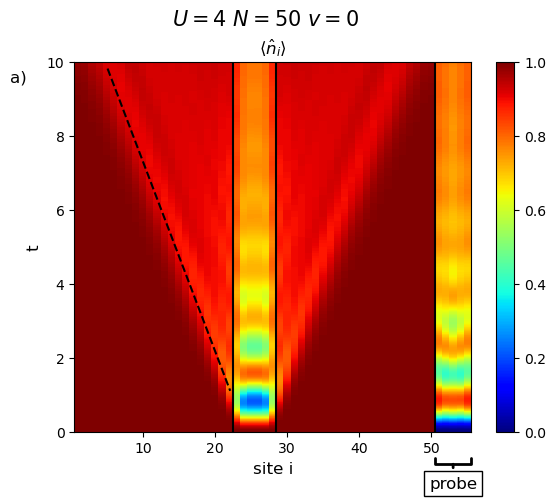}
	\includegraphics[width=\columnwidth]{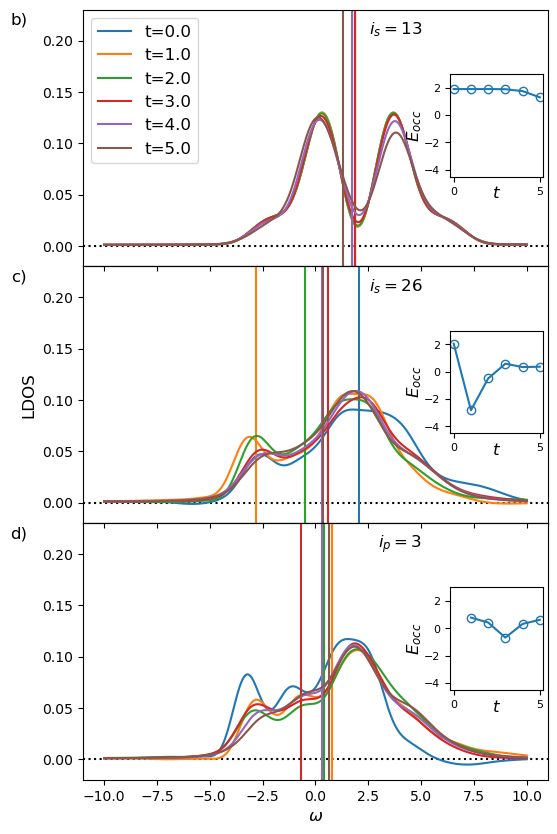}						
	\caption{
		MPS results for (a) the local particle density and the LDOS for (b) $i_s=13$, (c) $i_s=26$,
		and (d) $i_p=3$, plotted as in Fig.~\ref{fig:u0n26v0}, for $U=4$, half
		filling, i.e., $N=50$, and $v=0$.		
	}	\label{fig:u4n50v0}						
\end{figure}

At finite interaction,
$U = 4$, %
Fig.~\ref{fig:u4n50v0}(a) shows that the oscillations in the local
particle density between the sample and the nanoprobe are also present, but
are heavily damped.
The large damping is not unexpected, as 
the system is %
insulating for
$U = 4$ %
and half
filling, %
inhibiting charge transport.

In the LDOS at $i_s=13$, Fig.~\ref{fig:u4n50v0}(b), we observe two
peaks, which, at the beginning of the time evolution, have essentially
the same height. 
The splitting between the peaks is consistent with the size of the
Mott-Hubbard gap at this value of $U$, which, according to the Bethe ansatz, is $\Delta \approx 1.29$ ~\cite{Gebhard1997}.
However, note that the weight of the LDOS in Fig.~\ref{fig:u4n50v0} does not go to zero in the gap region. 
This is because of the limited resolution of our calculations; see Sec.~\ref{sec:methods}.
As in the previously discussed cases, %
the LDOS and $E_{\rm occ}$ on this site initially show only very weak time dependence, while at $t=5$ a clear change is visible. 
We associate this onset of change %
with the arrival of  the light cone, which induces a change of the number of particles on this site, and hence a change of the LDOS and $E_{\rm occ}$.

For $i_s=26$ and $i_p=3$, Figs.~\ref{fig:u4n50v0}(c) and (d),
we obtain a continuous LDOS with a smaller peak at energies below $E_{\rm occ}$ and a higher peak at energies larger than $E_{\rm occ}$.
On both sites, the LDOS changes in time, while, at later times, the changes become smaller. 
The overall structure of the LDOS for both sites and at all times is again comparable, as in the case of quarter filling. 
However, the gap is
no longer present %
after the coupling to the probe is turned on, indicating a melting of the Mott insulator in this region.

We expect that this melting of the Mott insulator will propagate %
through the system and that the gap in the LDOS will close with the arrival
of the light cone on the respective site. It remains an open question
as to %
whether the shape of the LDOS at longer times will approach the
results on site $i_s=26$ also further away from the nanoprobe.
Due to the high computational cost, %
we do not address %
this question %
further %
in this paper.

From the insets of Figs.~\ref{fig:u4n50v0}(b)-(d),
it can be seen %
that $E_{\rm occ}(t)$ behaves similarly to the previously discussed
cases, but that the damping is much stronger.
The behavior on sites $i_s=26$ and $i_p=3$ indicates that the
amplitudes of the oscillations are damped nearly completely %
to zero on the time scales that we have been able to treat;
in addition, the values of $E_{\rm occ}(t)$ at time $t=5$ are quite
similar, which corresponds to what we would expect if %
equilibration takes place.
Both aspects indicate that the LDOS reaches equilibrium on both
adjacent sites on the short time scales treated here. %
This is interesting because, %
in typical materials, the hopping strength is $t_h \sim 1$eV, which
translates to time scales ${\cal O}(1 \, \text{fs})$, %
corresponding to one time unit in Fig.~\ref{fig:u4n50v0}.
Thus, equilibration after coupling a Mott insulator to a nanoprobe
seems to happen on ultrafast time scales of a few femtoseconds.
To further investigate
the equilibration behavior of the Mott-insulating case, %
longer times and additional %
Mott-insulating systems would need to be treated,
which, due to the high computational expense, %
is left as a subject for future research.

\subsubsection{Interpretation of the results as a nonequilibrium LDOS}\label{subsec:discrepancies}

In the preceding %
discussion, we have tacitly interpreted our results as
if we were dealing with an equilibrium or near-equilibrium setup
(i.e., have applied linear response), so that %
the corresponding Fourier transform of the
Green's function, Eq.~\eqref{eq:LDOS}, can be rewritten in terms of a
Lehmann representation with explicitly positive weights. 
Out of %
equilibrium, %
the weights cannot be proven to be positive in general %
\cite{Kalthoff2}, so that, in principle, negative weights can
appear. %
This makes it difficult to interpret the corresponding %
results as %
a time-dependent LDOS. 
However, as can be seen in Figs.~\ref{fig:u0n26v0}-\ref{fig:u4n50v0},
all of the weights calculated here %
are positive, up to small artifacts, which may be due, at least in
part, to the way we compute the Fourier transform.
Hence, for the cases treated so far, it appears reasonable
to interpret the results as time-evolving  %
LDOS.

\begin{figure}[!]
	\includegraphics*[width=\columnwidth]{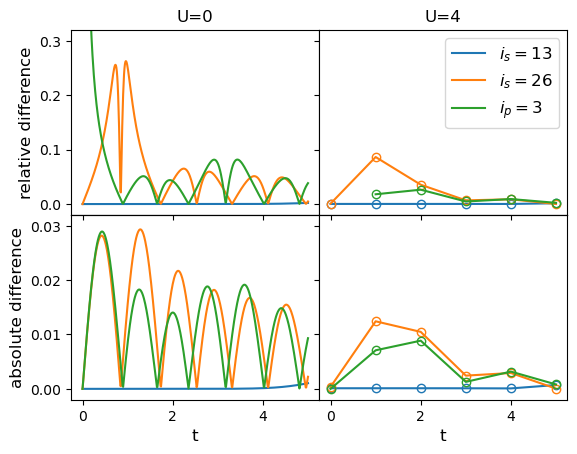}
	\caption{
		Absolute and relative discrepancy between the local
		expectation value of the local densities $\langle
		n_{\sigma,x}\rangle(t)$ and the integral over the
		lesser spectral function $\int d\omega \,
		B_{\sigma}^<(\omega,x)$. Left column: half filling,
		i.e., $N=50$, $U=0,\,v=0$. Right column: the same at
		$U = 4$.
		(Top row, relative difference; bottom row, absolute difference.)
		We define %
		the absolute difference here as $|\int d\omega \, B_{\sigma}^<(\omega,x)-\langle n_{\sigma,x}\rangle(t)|$ and the relative difference as $|\left(\int d\omega \, B_{\sigma}^<(\omega,x)\right)/\langle n_{\sigma,x}\rangle(t)-1|$.
		\label{fig:discrepancies}
	}
\end{figure}

This issue can be further investigated %
by studying the lesser and  greater
parts of the LDOS as defined in Eqs.~\eqref{eq:Bsmaller}
and~\eqref{eq:Bgreater}, respectively. 
In equilibrium, the expectation is that $\int d\omega \,
B_{\sigma}^<(\omega,x) = \langle n_{x,\sigma} \rangle$. 
Figure~\ref{fig:discrepancies} shows results for the discrepancy
between the two quantities %
in the out-of-equilibrium case for both the noninteracting and the
interacting,  %
$U=4$, cases %
at half filling with a resting %
nanoprobe. %
The agreement between $\int d\omega \, B_{\sigma}^<(\omega,x)$ and the
independently computed expectation values $\langle n_{x,\sigma}(t)
\rangle$ at later times is within a few percent, thus substantiating
our interpretation of $D_{\sigma}(\omega,x,t)$, as defined in
Eq.~\eqref{eq:LDOS}, as a good approximation to the time evolution of
the LDOS for the later times.
Interestingly, for the interacting system, the discrepancy is smaller
than for the noninteracting system.
At shorter times, $t \approx 1$, %
however, the discrepancy can become as large as %
$\sim 30\%$. 
This shows that here the system is %
in a strongly out-of-equilibrium %
regime in which %
the interpretation of our results as %
a time-dependent LDOS
must %
be treated with caution. %
Note that, in all cases, the absolute value of the discrepancy is very
small; at times $t \approx 1$, %
however, the small particle numbers lead to a large relative discrepancy.
The discrepancy in %
the actual %
values is larger at small times, and, again, %
is smaller for the interacting system.
 
For the case of a moving probe, similar behavior is obtained (see section \ref{subsec:v_0.55} and App.~\ref{app:v_one}). 
However, as we will see next,
the computed LDOS can, in addition, take on negative values, %
in particular, in the strongly out-of-equilibrium %
regime at short times. 

\subsection{Moving nanoprobe: $v>0$}\label{subsec:v_0.55}
Having explored the behavior when coupling a stationary nanoprobe to the sample at time $t=0$, we now treat cases in which the nanoprobe is moving at constant speed.
As in the stationary case, we 
also turn %
on the coupling between sample and a nanoprobe at the center of the
sample at time $t=0$; subsequently, the nanoprobe moves to the right
relative to the sample with speed $v$.

A natural scale for the velocities in the system is the velocity of
propagation of disturbances in the local particle density, i.e., the
speed associated with the previously discussed light cone,
$c\sim 2$. %
We treat %
two different velocities, $v=0.55$ ($\approx c/4$) and $v=1$ ($\approx c/2$).
Here we present the findings for $v=0.55$ in detail and complement the
discussion by highlighting the similarities and differences with the
case of the higher %
velocity of $v=1$, for which we present additional
results %
in App.~\ref{app:v_one}. %

\begin{figure}[!]
	\includegraphics[width=\columnwidth]{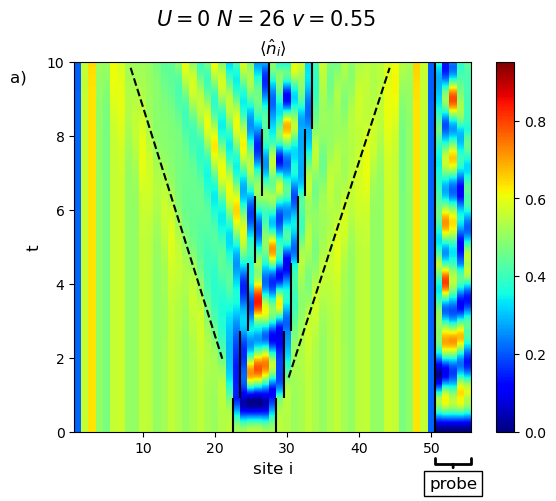}
	\includegraphics[width=\columnwidth]{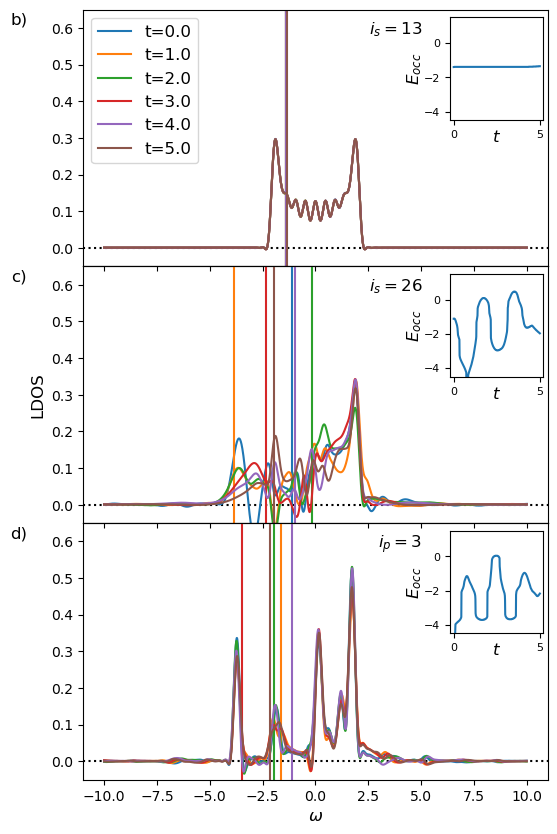}						
	\caption{%
		Exact results for (a) the local particle density and the LDOS for (b) $i_s=13$, (c) $i_s=26$,
		and (d) $i_p=3$, plotted as in Fig.~\ref{fig:u0n26v0},
                for $U=0$, quarter filling, i.e., $N=26$, and $v=0.55$.		
		}\label{fig:u0n26v0.55}						
\end{figure}

In the following, we will use the resting case, $v=0$, as a reference, 
highlighting %
the similarities and differences of the behavior in the two moving cases.
For all velocities 
$v>0$, we find that there are %
stronger wave fronts to the left of the
right-moving nanoprobe and weaker ones to its right; see, e.g.,
Fig.~\ref{fig:u0n26v0.55}(a) for $v=0.55$, $U=0$, and
quarter filling.
Note that the slope of the wave fronts is still consistent with a
constant speed $c\approx 1.6$ for the light cone. 
The effect
of the movement of the probe %
is that %
particles ``leak out'' %
of the nanoprobe in the wake
of its motion. 
The %
speed of $v=0.55$ is an %
interesting case
because, for this speed,  we find that %
these ``wake'' effects %
are particularly
pronounced, %
in contrast to the $v=1$ case (see App.~\ref{app:v_one}).
For $v= 0.55$, %
the motion of the probe is spatiotemporally commensurate with
the oscillations of the particle density, which leads to constructive
interference to the left of the moving probe.
Therefore, the wave fronts are
very pronounced for this particular velocity, as can be seen
in Fig.~\ref{fig:u0n26v0.55}(a).
Strong oscillations in the local particle density
between the nanoprobe and the system are still present,
but, for the moving nanoprobe, they are mostly concentrated at its trailing
edge rather than its center as in the stationary case.

The LDOS for this case is displayed in
Figs.~\ref{fig:u0n26v0.55}(b)--(d), for the same three sites as
before, $i_s=13$, $i_s=26$, and $i_p=3$.
Note that the LDOS is now, in general, time dependent, also for %
$U=0$.
As at $v=0$, the LDOS at site $i_s=13$ is essentially
time independent due to the fact that %
the nanoprobe is moving away from this site so that,
at the times treated, the wake caused by the moving nanoprobe
has not yet reached it. %
At site $i_s=26$, Fig.~\ref{fig:u0n26v0.55}(c), the evolution of the
LDOS differs significantly from that of the resting case,
Fig.~\ref{fig:u0n26v0}(c). %
One marked feature is that %
the results can now take on substantially negative values,
in particular, for times $t<3$.
However, %
at later times, the discrepancy between $\langle
n_{i,\sigma}\rangle$ and $\int d\omega \, B^<_{\sigma}(\omega)$ is
only a few percent, a behavior similar to that of the case of the
stationary nanoprobe.
Note that $i_s=26$ is a stationary point
in the center of the sample %
and therefore becomes decoupled from the nanoprobe at time $\approx6.4$, which is
shorter than the time interval over which we perform the
Fourier transform from $t$ to $\omega$.
For the faster moving case, $v=1$ 
(see App.~\ref{app:v_one}), site $i_s=26$ %
already decouples %
from the nanoprobe at times $t \approx 3.5$, so that the LDOS,
Fig.~\ref{fig:u0n26v1}(b),
takes on, approximately, the decoupled equilibrium form, i.e., that of
Fig.\ \ref{fig:u0n26v0}(b), for $t\geq4$. 

The LDOS at $i_p=3$, Fig.~\ref{fig:u0n26v0.55}(d), is %
essentially
time independent, despite the fast motion of the
nanoprobe,
a behavior also found in the $v=1$ case; see App.~\ref{app:v_one}.
The structure and the positions of the peaks are
roughly %
the same as for the resting %
nanoprobe, Fig.~\ref{fig:u4n26v0}(d).
Regarding $E_{\rm occ}(t)$, we see a similar behavior
to that of the %
resting case for $i_s=13$ and $i_p=3$.
However, $E_{\rm occ}(t)$ appears to increase with time for
$i_s=26$, unlike in the resting case.

\begin{figure}[!]
	\includegraphics[width=\columnwidth]{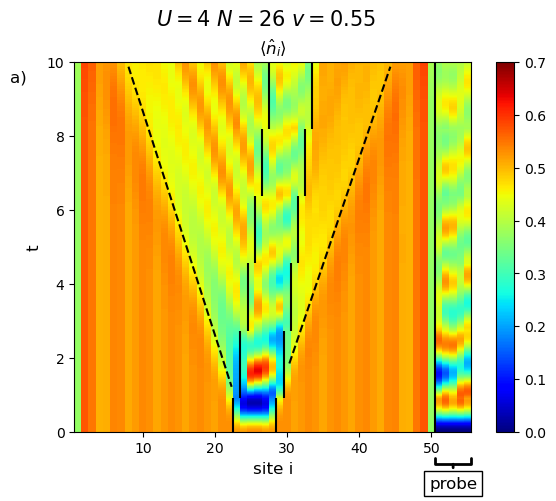}
	\includegraphics[width=\columnwidth]{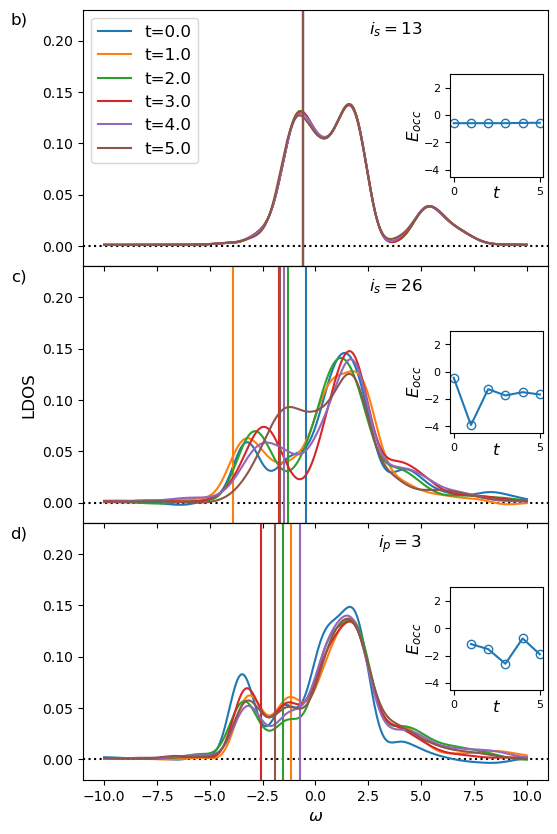}							
	\caption{
		MPS results for (a) the local particle density and the LDOS for (b) $i_s=13$, (c) $i_s=26$,
		and (d) $i_p=3$, plotted as in Fig.~\ref{fig:u0n26v0},
                for $U=4$, quarter filling, i.e., $N=26$, and $v=0.55$.	
		}	\label{fig:u4n26v0.55}						
\end{figure}

Turning on the interaction to $U=4$ in the quarter-filled case with
$v=0.55$, Fig.~\ref{fig:u4n26v0.55}, we see that 
$\langle n_i(t)\rangle$ 
has similar features to that of the
$U=0$ case, Fig.~\ref{fig:u0n26v0.55}(a).
Note, however, that now the oscillation is significantly
more damped.
As before, the LDOS at site $i_s=13$, %
Fig.~\ref{fig:u4n26v0.55}(b), is approximately time independent since the light cone has not reached this position on the time scales treated by us.
Comparing the LDOS on site $i_s=26$ to the resting case, $v=0$, we
observe %
that the overall picture is comparable, but that there are now
more changes in time in the relevant energy region; notably, at $t=3$,
a dip evolves between two peaks at $\omega \approx -2.5$ and $1$.
However, this dip disappears at later times.
It is notable that---in contrast to the
noninteracting case---the LDOS
takes on %
(up to minimal effects) only positive values for all values of
$\omega$ shown.
As discussed in Section ~\ref{subsec:discrepancies}, the
discrepancy of the occupations is only a few percent, so that, again,
interpreting the spectrum as a quasi-equilibrium LDOS seems to be justified. 
For $i_p=3$, Fig.~\ref{fig:u4n26v0.55}(d), there is only weak time
dependence.
The results are very similar to the ones for $v=0$, with only minor
qualitative differences.
This is interesting, as %
it indicates that the motion of the nanoprobe seems to  not
significantly affect the time evolution of the LDOS on its
sites.
This similarity of the
  LDOS on site $i_p=3$ to the $v=0$ case remains present when the
  nanoprobe velocity is higher, $v=1$;  see App.~\ref{app:v_one}.

The time-dependent behavior of $E_{\rm occ}(t)$ 
for $v=0.55$, depicted in the insets of
  Figs.~\ref{fig:u4n26v0.55}(b)-(d),
is very 
similar to that of the resting case, Figs.~\ref{fig:u4n26v0}(b)-(d),
However, for $i_s=26$, $E_{\rm occ}(t)$ appears to settle
to a stationary value within the
simulation time, in contrast to the $v=0$ case.
This difference in behavior is due to the fact that the probe moves away
from this position so that local observables
equilibrate faster than
when the probe is resting.
At site $i_p=3$, however, $E_{\rm occ}(t)$ keeps changing in time,
as might be expected because the nanoprobe continues to move over the sample.

\begin{figure}[!]
	\includegraphics[width=0.95\columnwidth]{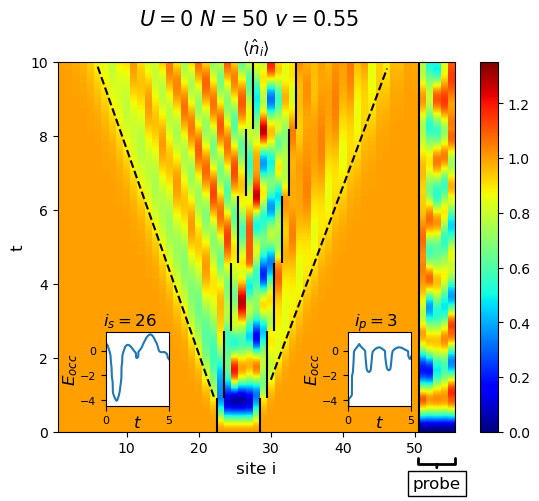}							
	\caption{
	Exact results for the local particle density, plotted as in
        Fig.~\ref{fig:u0n26v0}, for $U=0$, half filling, i.e., $N=50$, and $v=0.55$. The occupation energy
	$E_{\rm occ}(t)$ is shown as an inset for sites $i_s=26$ and $i_p=3$. The LDOS in this case is not explicitly shown, as it is the same as in Fig.~\ref{fig:u0n26v0.55}.
	}
        \label{fig:u0n50v0.55}
\end{figure}

We continue examining probe velocity
  $v=0.55$, but now increase the initial
band filling to half filling, first taking $U=0$.
Figure~\ref{fig:u0n50v0.55} displays the $U=0$ local particle density.
As can be seen by comparing with
  Fig.~\ref{fig:u0n26v0.55}(a),
the wave fronts to the left of the nanoprobe are even more
pronounced than in the quarter-filled case.
Furthermore, there is an even stronger build-up of local particle
density 
inside the probe.
At larger times, %
the local particle density rapidly changes over to being concentrated
at the leading edge of the nanoprobe.
Note that the build-up of local density is even stronger in
  the $v=1$ case,  with it being 
concentrated at %
the trailing edge of the probe for %
times up to $t\approx7$; see App.~\ref{app:v_one}.

\begin{figure}[!]
	\includegraphics[width=\columnwidth]{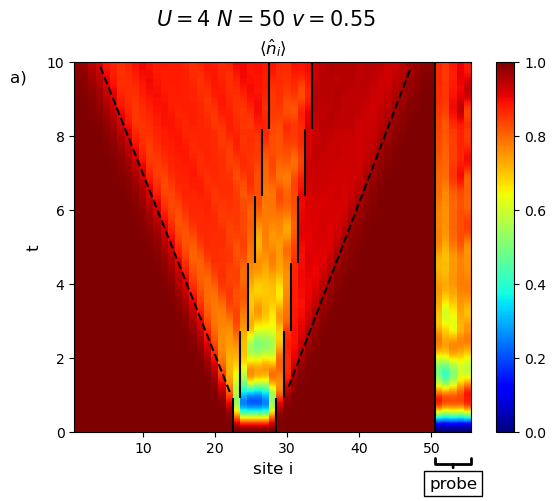}
	\includegraphics[width=\columnwidth]{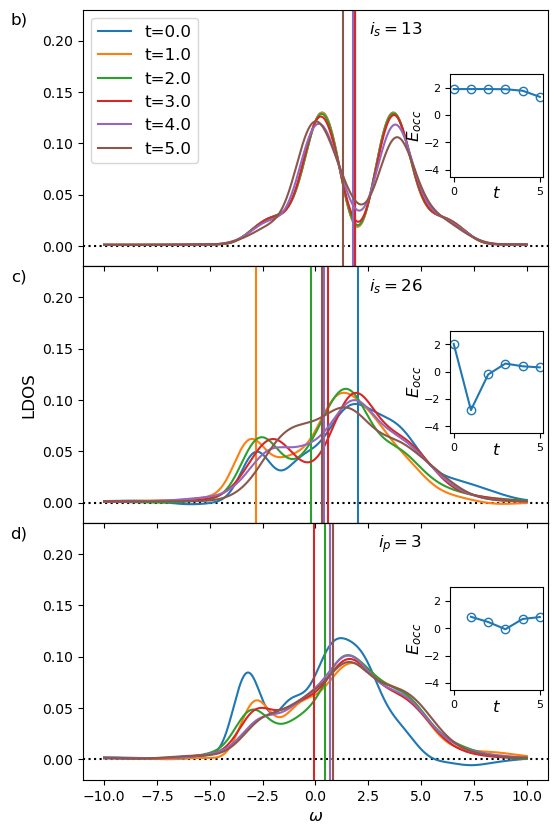}						
	\caption{
		MPS results for (a) the local particle density and the LDOS for (b) $i_s=13$, (c) $i_s=26$,
		and (d) $i_p=3$, plotted as in Fig.~\ref{fig:u0n26v0},
		for $U=4$, half filling, i.e., $N=50$, and $v=0.55$.
	}	\label{fig:u4n50v0.55}						
\end{figure}

We now consider the interacting case, $U=4$,
  Fig.~\ref{fig:u4n50v0.55}, remaining at half filling and $v=0.55$.
As in the half-filled, interacting, $v=0$ case,
  Fig.~\ref{fig:u4n50v0}(a),
we see once again %
that the oscillation between the nanoprobe %
and the
system is heavily damped.
We notice that there is a lower
particle density towards the left of the probe inside the light cone,
consistent with the probe moving to the right.
  For higher probe velocity, $v=1$ (see App.~\ref{app:v_one}),
  the local particle density behaves similarly, except that the
  oscillations in particle density in the sample are less strong, as
  the probe moves more quickly over the sample.

Examining the LDOS, first at site $i_s=13$,
Fig.~\ref{fig:u4n50v0.55}(b), we can again see the peaks at
$\omega\approx0$ and $4$ for $i_s=13$ and a gap at
$\omega\approx2$, similar to the $v=0$ case,
Fig.~\ref{fig:u4n50v0}(b).
The movement of the probe causes the peak at $\omega\approx4$ to
shrink slightly for later times $t$,
 a feature also found at higher probe velocity $v=1$
  (see App.~\ref{app:v_one}). %
In the LDOS at site $i_s=26$, Fig.~\ref{fig:u4n50v0.55}(c), one can see
that the Mott gap in the LDOS again disappears due to coupling to the
empty nanoprobe.
The LDOS %
features peaks at $\omega\approx-2.5$ and $2$, and
only changes slightly with time.
At the latest time that we have reached, $t=5$, the LDOS does show small
changes, building up spectral weight between the two peaks.
This is most likely due to the
nanoprobe becoming decoupled from site $i_s=26$ at $t\approx6.4$, which is
within the time interval over which the Fourier transform is carried out.
This effect is not yet noticeable at $t=4$, probably because
the Hann window that we apply weights times closer to the lower bound
of the integral more strongly.
At the initial time, $t=0$,
the  LDOS at probe site $i_p=3$, Fig.~\ref{fig:u4n50v0.55}(d), exhibits peaks at
$\omega\approx-4$ and $2$.
For $t\geq1$, both peaks diminish slightly, and the LDOS is shifted a
small amount towards larger $\omega$ values.
The LDOS then appears to settle in to %
this slightly altered shape for
all times $t\geq1$.
Interestingly, this form closely resembles that of the $i_s=26$ LDOS
[\textit{cf.}\ Fig.~\ref{fig:u4n50v0.55}(c)], 
which seems reasonable because %
the two sites are directly connected via a
hopping term.

Examining the behavior of $E_{\rm occ}(t)$
[insets of Figs.~\ref{fig:u4n50v0.55}(b), (c), and (d)],
it appears to be heavily damped in time, similarly as in the stationary ($v=0$)
half-filled system, Figs.~\ref{fig:u4n50v0}(b), (c), and (d).
For $i_s=26$ and $i_p=3$, $E_{\rm occ}(t)$ seems to settle to
approximately the same value, which is slightly
larger than zero.

\section{Discussion and Summary}
\label{sec:summary}
In this paper, we have investigated the coupling of an empty
nanoprobe to a one-dimensional system of correlated electrons for 
two different fillings, half filling and quarter filling, treating the
case of a stationary nanoprobe as well as that of a moving one.
We have taken both the sample and the nanoprobe to have intermediate local
Coulomb interaction strength, $U=4$, and have contrasted the behavior
with that of the noninteracting, $U=0$, case.
In all scenarios that we have studied, the coupling between the sample and
the nanoprobe, consisting of nearest-neighbor hopping terms
between proximate lattice sites of the sample and corresponding sites
of the nanoprobe, is suddenly turned on.
Using an MPS formulation of the DMRG and treating the time dependence
using an TDVP scheme, we have studied the time-dependent behavior
of the local particle density $\langle n_i \rangle(t)$ and of the LDOS
$D_\sigma(t,\omega,x)$, as defined by Eq.~\eqref{eq:LDOS}.

As a reference system, we take the noninteracting case with a stationary
nanoprobe.
Turning on the hopping between the sample and the
empty nanoprobe at $t=0$ induces long-lived oscillations in the
local particle density  $\langle n_i \rangle(t)$ in the sites in the
probe %
and in the sample in the vicinity of the nanoprobe. %
As can be shown rigorously, the LDOS for this system is time independent.
However, the occupation energy $E_{\rm occ}(t)$,  defined by
Eq.~\eqref{eq:E_occ}, which is a measure of
the local chemical potential, oscillates in time, with opposing phases
of oscillation on the site on the nanoprobe and the proximate site on
the sample due to the strong tunneling between them, reflecting the
oscillations in the local particle density.

For systems with finite interaction strength, $U=4$,
oscillations in the local particle density are still present, but
become more strongly damped, especially at half filling.
At both quarter and half filling, the LDOS is no longer 
time independent, but does evolve towards a stationary form at longer
times; this holds for both stationary and moving nanoprobes.
The oscillations in $E_{\rm occ}(t)$ are, in general, still present,
but are more strongly damped than for $U=0$, indicating that the
system seems to already attain local equilibrium on the short
time scales that we have been able to treat here.
This damping is somewhat stronger for the
half-filled case than for the quarter-filled one, %
reflecting
the reduced charge mobility in the initially Mott insulating phase,
consistent with the stronger damping in the local particle
density.

For the initially half-filled sample,
the gap in the LDOS
in the sample site situated at the center of the nanoprobe closes
immediately upon suddenly coupling the sample and probe, indicating
the local breakdown of the Mott insulator.
However, for sample sites further away from the nanoprobe, some time
is required before the perturbation induced by coupling the two
subsystems reaches the observation position.
In this way, one observes a ``melting'' of the Mott
insulator that propagates through the system with time.
For the quarter-filled sample, the LDOS is clearly metallic
far from the nanoprobe, where it remains stationary, in the
sample under the nanoprobe, where there is moderate change with
time, and in the nanoprobe, where there are only small changes with
time once the connection between sample and probe is turned on.

For a moving nanoprobe, the local environment in the sample,
in particular, the local particle density, changes as the nanoprobe
moves over it.
Note, however, that changes in the sample propagate outwards with
a velocity given by the light cone, which is significantly larger than
the speed of movement of the nanoprobe
and can be clearly seen in the time and space dependence
of the local particle density.
The primary effect of the movement of the nanoprobe is to reduce
backflow effects into the sample, as the back-tunneling from the probe
to the sample is spread over different sample sites as the probe
moves.
The behavior of the LDOS is similar to that of the case of a resting nanoprobe.
Interestingly, on the sites of the moving probe, the LDOS behaves 
essentially as if the probe were at rest.
On the sites in the sample,
the behavior of the LDOS is similar to that in the stationary
case, up to times at which the site initially under the nanoprobe is
no longer under the influence of the nanoprobe because it moves far
enough away.

These features of the cases with a moving nanoprobe remain
qualitatively the same when the speed of the nanoprobe is increased.
The exceptions are that for the slower nanoprobe velocity, $v=0.55$,
there are interference effects in the local particle density due to
commensuration of spatial and temporal oscillations of the local
particle density that are not present at the higher nanoprobe speed,
$v=1$.
In addition, the decoupling of the nanoprobe from sample sites
initially under the nanoprobe occurs on shorter time scales for the
fast nanoprobe velocity, so that quasi-equilibration of the LDOS
occurs sooner on such sites.

We remark that the nonequilibrium LDOS that we have computed
[Eq.~\eqref{eq:LDOS}] is not restricted to be strictly positive
semi-definite.
Negative values are incompatible with the interpretation as spectral
weights that one is used to for an equilibrium LDOS.
However, we, in fact, find that negative weights only occur in very
few cases (most markedly, %
in the $U=0$ case with a
moving nanoprobe), and, for these cases, only at short times.
In these cases, we also find that there is a discrepancy between
the local particle density directly calculated as 
$\langle n_{i,\sigma}\rangle(t)$ and that obtained from
$\int d\omega \, B^<_{\sigma}(\omega)$.
We characterize this anomalous short-time behavior as being
strongly nonequilibrium.   
Thus, the general picture is that the discrepancies in the local
particle density and the appearance of negative values of the LDOS
disappear at later times (especially in the interacting system),
which is consistent with reaching an 
equilibrium state, so that a description in terms of standard linear
response theory is applicable. 
It is interesting to see that this happens on the very short time
scales treated by us; for a typical material, the hopping strength
$t_h \sim 0.5-1\, \text{eV}$.
Since the units of our time scale are $\hbar/t_h$, this implies
that our calculations typically reach
$\sim 5-10\, \text{fs}$, which lies in the ultrafast regime of even
highly time-resolved experimental techniques such as
pump-probe experiments.
Developing such time-resolved local spectroscopy with STM is an
ongoing challenge~\cite{Review_timedependent_STM}. However, one can envisage that
similar scenarios to the ones proposed here
could be studied using single-site microscopes in
experiments
on optical lattices,
which realize the Hubbard model~\cite{esslinger_review,single_site_microscope1,single_site_microscope2,single_site_microscope3,single_site_microscope4}.  
In either case, it will be interesting to study how the states
developing on the ultrashort time scales treated by us
here affect the
behavior at the later times accessible to these experiments. %

It would %
therefore be interesting to also study the time evolution of the LDOS and
other dynamical quantities for other strongly correlated
systems.
For example, investigating a similar scenario to the one
treated here
in correlated charge density wave (CDW) insulators
could allow one to study if and how the melting of a CDW state
propagates through the system after a local perturbation. 
Since topological phases are either protected by local symmetries
(e.g., symmetry-protected topological phases %
in one dimension) or by long-range entanglement (in particular, in two-dimensional systems), it
would be
interesting to study the interplay of the local perturbation with
these protection mechanisms by studying the time evolution of the
LDOS.

\begin{acknowledgments}
We thank Martin Wenderoth for helpful
discussions. S.R.M. acknowledges financial support by
Deutsche Forschungsgemeinschaft Grant No. 217133147/SFB
1073, Project No. B03. This paper was initiated during
a visiting professorship of S.R.M. at
Philipps-Universität Marburg from 2020 to 2021.
\end{acknowledgments}

\appendix
\section{Numerical Accuracy}
\label{subsec:convergence}
\begin{figure}[!]	
  \includegraphics[width=\columnwidth]{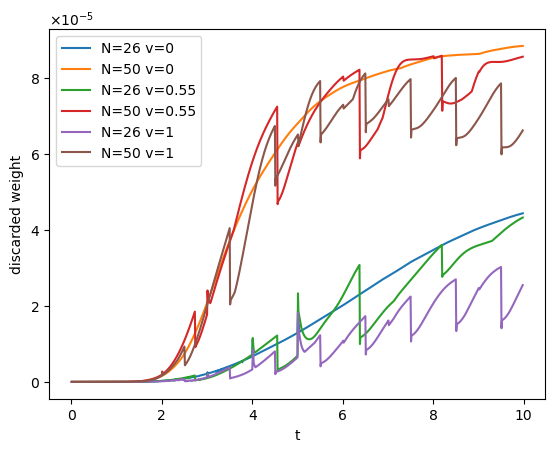}							
  \caption{Maximum discarded weight $\delta$ during a TDVP
    time step as a function of %
    the evolution time $t$ for $U=4$, with $L=50$ sample sites and
    $L_p=5$ nanoprobe sites and initial filling and probe speed as
    indicated in the legend.
  }\label{fig:discw}						
\end{figure}

The truncation error of the time evolution
has been found to be proportional to the
square root of the discarded weight: $\epsilon\approx\sqrt{\delta}$
\cite{Paeckel2019}.
In Fig.~\ref{fig:discw}, we find %
that,  for the half-filled system, 
the discarded weight for
later times 
reaches just under
$\delta\approx10^{-4}$, resulting %
in an error of
$\approx10^{-2}$.
This level of accuracy should be sufficient to discern features of 
  the LDOS that dominate its qualitative behavior.
Furthermore, %
as discussed in Sec.~\ref{subsec:MPSmethods}, 
this level of error is also
overshadowed by the %
limited resolution of our calculation of the LDOS
caused by the fact that we are only
able to
carry out the integrals in Eqs.~\eqref{eq:Bsmaller} and
\eqref{eq:Bgreater}  to  $\tau_{\text{max}}=5$.

\begin{figure}[!]	
  \includegraphics[width=\columnwidth]{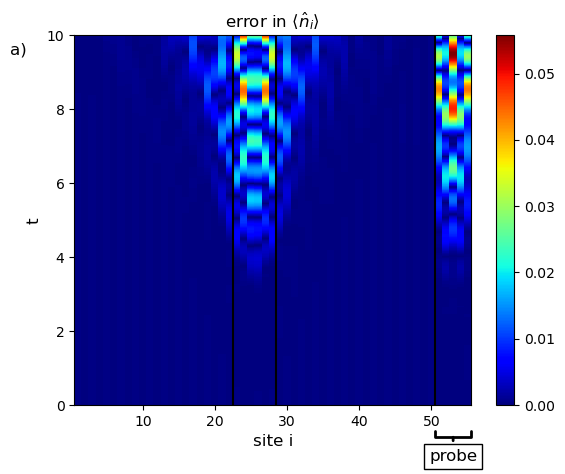}			
  \includegraphics[width=\columnwidth]{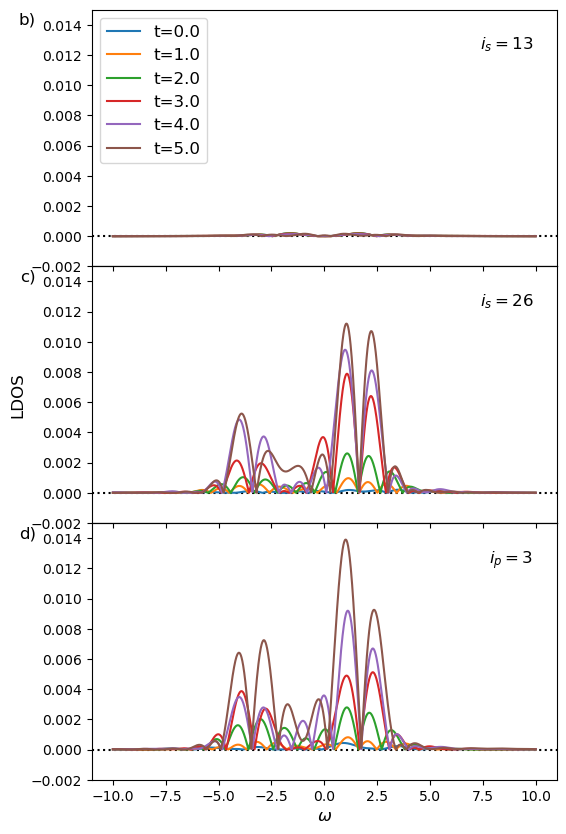}							
  \caption{ %
    Results of the TDVP time evolution, with $L=50$ sample sites and
    $L_p=5$ nanoprobe sites, compared to exact
    diagonalization for $U=0$ and probe velocity $v=0$ at half filling,
    i.e., $N=50$. 
    Here (a) depicts the absolute difference in the particle density
    $\langle n_i \rangle$, and (b)-(d) depict the absolute
    difference in the LDOS at the indicated sites.
    Note that the
    maximum integration time  %
    $\tau_{\text{max}}=5$ for both the
    exact diagonalization and TDVP calculations in order to make
    the results comparable.
}\label{fig:LDOS_error_v0}						
\end{figure}

In Fig.~\ref{fig:LDOS_error_v0}(a), it can be seen that the
error in the local particle density \erwartung{n_i} becomes
appreciable at times after $t\approx 4$ in and in the vicinity of the
nanoprobe and continues to grow and spread out as time progresses.
The error is thus concentrated in regions in which the most change in
time takes place.
The LDOS far away from the nanoprobe, $i_s=13$, 
Fig.~\ref{fig:LDOS_error_v0}(b), shows no significant error, whereas
that under and in the nanoprobe, Figs.~\ref{fig:LDOS_error_v0}(c) and ~\ref{fig:LDOS_error_v0}(d),
shows absolute errors that grow with time, becoming significant at
later times.
Nevertheless, we consider %
them to be small enough to ensure that the
qualitative behavior of the LDOS,
which is limited anyway due to taking the maximum time in our Fourier
integration,
Eqs.~\eqref{eq:Bsmaller} and ~\eqref{eq:Bgreater},
to be $\tau_{\text{max}}=5$, remains unchanged.
The maximum absolute difference in the LDOS at sites $i_s=26$ and
$i_p=3$ for later times is about $0.014$.
The largest peaks that are resolved %
in the LDOS (for $\tau_{\text{max}}=5$)
are approximately $0.15$ in amplitude.
This is an error of about 10\% %
for the largest peaks and only
occurs at later times.
Despite this appreciable error, the accuracy should be sufficient to
resolve the larger peaks in the LDOS.
When evaluating the presence and size of the smallest peaks, however,
which reach from approximately $0.015$ to $0.05$, one
cannot necessarily make a definite determination. %
Nevertheless, we emphasize that the errors at smaller times and for lattice site
$i_s=13$ are almost negligible and yield more than sufficiently
accurate results.
Further details on the numerical
accuracy of calculations for the interacting system, in particular,
for the moving probe with $v=1$ and with regard to the
bond-dimension dependence, can be found in the
supplemental material~\cite{supplemental_material}.

\section{Results for $v=1$}
\label{app:v_one}
\begin{figure}[!]
	\includegraphics[width=\columnwidth]{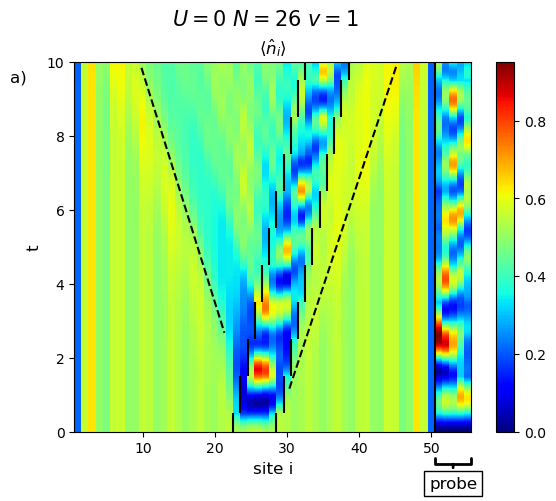}
	\includegraphics[width=\columnwidth]{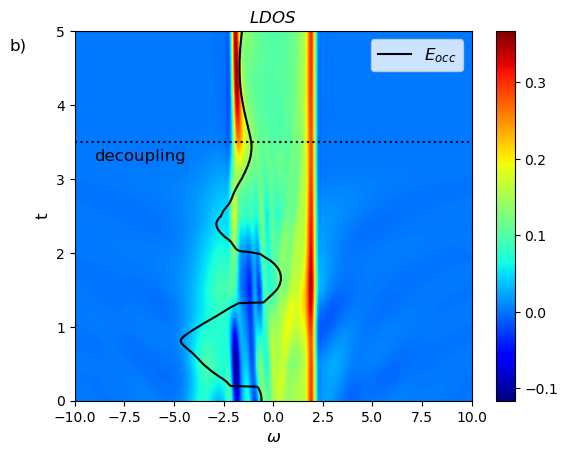}						
	\caption{Exact results for the nanoprobe model, with $L=50$
          sample sites and $L_p=5$ nanoprobe sites, for the
		system initially at quarter filling, i.e., $N=26$, with
		$U=0$ %
		and nanoprobe %
		velocity $v=1$.
		Here
		(a) depicts the expectation value of the local particle
		density $\langle {n}_i \rangle$ as a color density
		plot as a function of lattice site $i$ and time $t$,
		and 
		(b) plots the LDOS %
		at site $i_s=26$ as a function of the frequency $\omega$.		
		The occupation energy
		$E_{\rm occ}(t)$ at site $i_s=26$ is shown as a solid line.
		The dotted line indicates the decoupling of the probe from lattice site $i_s=26$ at $t=3.5$.
	}\label{fig:u0n26v1}							
\end{figure}

In this appendix, we
present results for a probe velocity of
$v=1$ for the same sets of remaining %
parameters as the results for $v=0.55$
presented in Sec.~\ref{subsec:v_0.55}.
Note that %
most of the features of the $v=1$ case are very similar to those
of the $v=0.55$ case;
many of the important
differences and similarities have already been highlighted in
Sec.~\ref{subsec:v_0.55}.
In the following, we will present $v=1$ results for both quarter and
half filling for both $U=0$ and $4$ and
briefly describe salient aspects.
We first treat the case of %
zero interaction strength, $U=0$, and quarter filling.
Comparing the local particle density, Fig.~\ref{fig:u0n26v1}(a), with
that for the $v=0.55$ case,  Fig.~\ref{fig:u0n26v0.55}(a),
we observe the same qualitative behavior,
albeit with more pronounced wave fronts to the left of the nanoprobe
for $v=0.55$, as discussed in Sec.~\ref{subsec:v_0.55}.
Comparing the LDOS for sample site $i_s=26$,
Fig.~\ref{fig:u0n26v1}(b), to that of the $v=0.55$ case, Fig.~\ref{fig:u0n26v0.55}(c),
we can see that the right peak in the LDOS is preserved throughout the
whole time evolution.
However, the left peak in
Fig.~\ref{fig:u0n26v0.55}(b) is lost after the coupling of the probe
to the system is turned on at $t=0$.
Note, however, that this peak is restored at
$t\approx3.5$  for $v=1$, as can be seen in Fig.~\ref{fig:u0n26v1}(b).
This is due to the fact that, at this point in time,
the probe has traveled far enough across the system that site $i_s=26$
is decoupled from the probe.

\begin{figure}[!]
	\includegraphics[width=\columnwidth]{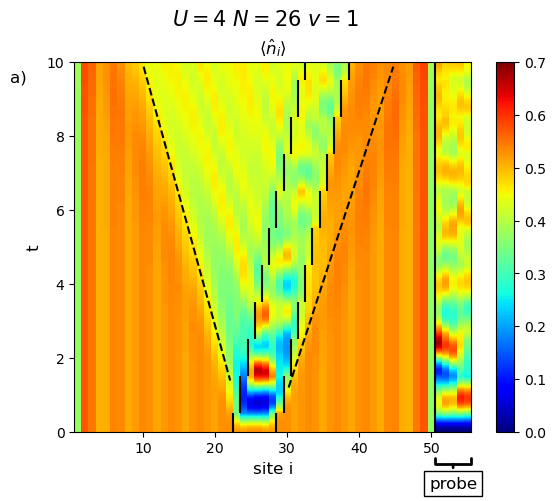}
	\includegraphics[width=\columnwidth]{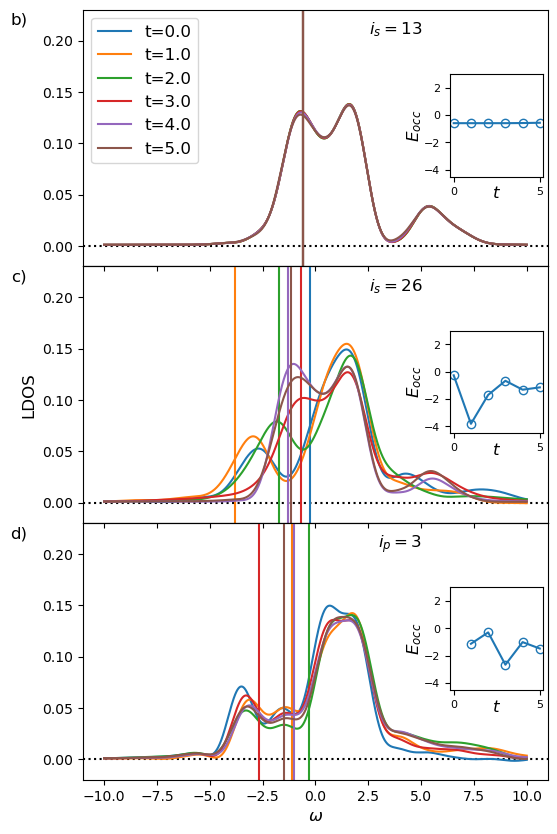}						
	\caption{
		MPS results for (a) the local particle density and the LDOS for (b) $i_s=13$, (c) $i_s=26$,
		and (d) $i_p=3$, plotted as in Fig.~\ref{fig:u0n26v0},
                for $U=4$, quarter filling, i.e., $N=26$, and $v=1$.
	}		\label{fig:u4n26v1}					
\end{figure}

We now turn to the case of intermediate interaction strength, $U=4$,
for the quarter-filled system.
Comparing the local particle density,
Fig.~\ref{fig:u4n26v1}(a),
with that for the $v=0.55$ case, Fig.~\ref{fig:u4n26v0.55}(a), we
again see mostly the same behavior except that there are more
pronounced wave fronts to the left of the nanoprobe.
For the LDOS, comparing Figs.~\ref{fig:u4n26v1} (b)-(d) with Figs.~\ref{fig:u4n26v0.55}
(b)-(d), we can see that the LDOS at sites $i_s=13$ and $i_p=3$ are almost
unaffected by the change in speed of the probe.
In Fig.~\ref{fig:u4n26v1} (c),
we again observe that the left peak in the LDOS is
restored for times $t\geq4$ for the probe speed of $v=1$ due to
decoupling of the nanoprobe from site $i_s=26$.

\begin{figure}[!]
	\includegraphics[width=\columnwidth]{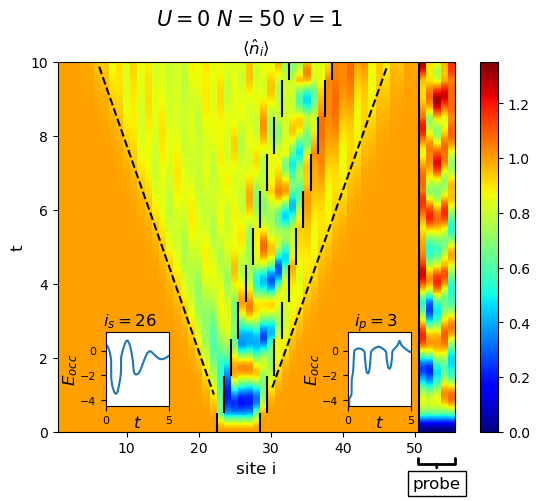}								
	\caption{
		Exact results for the local particle density, plotted
                as in Fig.~\ref{fig:u0n26v0}, for $U=0$, half filling,
                i.e., $N=50$, and $v=1$. The occupation energy
		$E_{\rm occ}(t)$ is shown as an inset for sites $i_s=26$ and $i_p=3$. The LDOS in this case is not explicitly shown, as it is the same as in Fig.~\ref{fig:u0n26v1}.}\label{fig:u0n50v1}							
\end{figure}

For the $U=0$ case in the half-filled system,
comparing Fig.~\ref{fig:u0n50v1}
with
Fig.~\ref{fig:u0n50v0.55}, %
the behavior is qualitatively the same
except for the more pronounced wave fronts for $v=0.55$, also seen at
quarter filling.
One can see that there are more particles collected in the nanoprobe
over time for
$v=1$; this is due to the fact that the higher probe speed makes more particles
available to tunnel into the probe.

\begin{figure}[!]
	\includegraphics[width=\columnwidth]{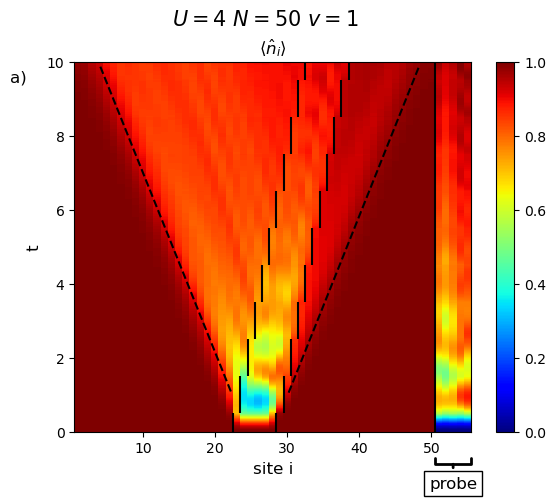}
	\includegraphics[width=\columnwidth]{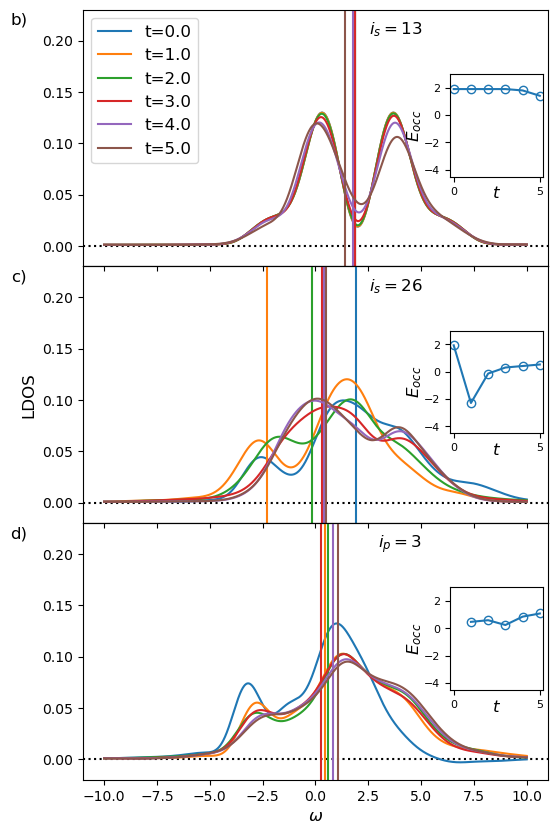}					
	\caption{
		MPS results for (a) the local particle density and the LDOS for (b) $i_s=13$, (c) $i_s=26$,
		and (d) $i_p=3$, plotted as in Fig.~\ref{fig:u0n26v0},
                for $U=4$, half filling, i.e., $N=50$, and $v=1$.
	}	\label{fig:u4n50v1}						
\end{figure}

Finally, for interaction strength $U=4$ and half filling,
comparing Fig.~\ref{fig:u4n50v1} with the $v=0.55$ case,
Fig.~\ref{fig:u4n50v0.55},
the local particle density, Fig.~\ref{fig:u4n50v1}(a),
behaves similarly for the two probe speeds, but
again, more particles are collected in the probe over
time for $v=1$.
In the LDOS,
the behavior for site $i_s=13$, Fig.~\ref{fig:u4n50v1}(b),
is essentially unaffected by the probe speed, showing a double-peak
structure characteristic of a Mott insulator where there is only a
slight shift in the relative weight of the two peaks with time.
For site $i_s=26$,  the behavior at $v=1$, Fig.~\ref{fig:u4n50v1}(c), is similar
to that for $v=0.55$, Fig.~\ref{fig:u4n50v0.55}(c), in that the
two-peak structure
characterizing the Mott insulator %
immediately disappears at $t=0$, and two weaker
peaks at small times evolve into a structure characterized by one
peak.
In contrast to $v=0.55$, however, a new two-peak structure with a
stronger, broader, left peak emerges and remains stationary at the two longest times,
$t=4.0$ and $5.0$.
Within the nanoprobe at $i_p=3$, the behavior for $v=1$,
Fig.~\ref{fig:u4n50v1}(d), and $v=0.55$, Fig.~\ref{fig:u4n50v0.55}(c),
is %
virtually identical, showing a two-peak structure without a
discernible gap at small times washing 
out into a broad structure with vestigial peaks at larger times.

\section{Charge velocity}
\label{app:u_rho}

\begin{figure}[!]
  \includegraphics[width=\columnwidth]{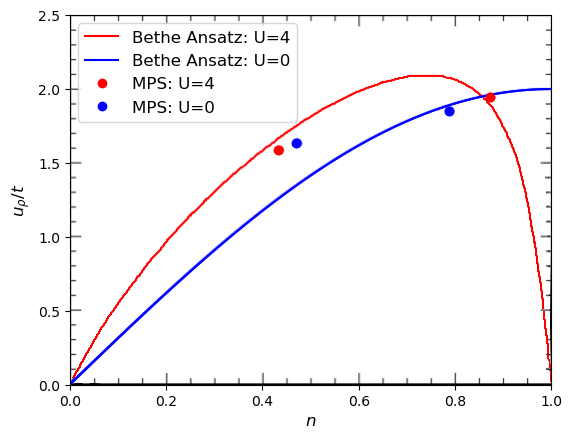}					
  \caption{
    Calculated values for the charge velocity
      (solid lines)
      as well as light cone velocities measured from our numerical
      (round points) for nanoprobe velocity $v=0$.
      For $U=0$, we take $u_\rho=v_F=2t\sin\left(\pi n\right)$, the
      Fermi velocity, as the charge velocity, whereas for $U=4$, we
      plot a curve taken from Bethe-ansatz calculations of
      Ref.~\onlinecite{schulz1993} (in particular, Fig.~12).
      The light-cone velocity is obtained from 
      fits of the wavefront in the local particle densities of
      Figs.~\ref{fig:u0n26v0}(a), \ref{fig:u4n26v0}(a), \ref{fig:u0n50v0}, and
      \ref{fig:u4n50v0}(a) above. 
      For these data, we take the value of the filling $n$ on the
      abscissa to be the average of the particle density within the first
      trough in the wavefront.}
  \label{fig:chargevel}
\end{figure}

In this appendix, we elucidate theoretical estimates of light-cone
velocity and compare with fitted values from the local particle
densities for a nanoprobe velocity $v=0$ and interaction strength
$U=0$ and $U=4$.
As an estimate of the light-cone velocity, we take the velocity
$u_\rho$ of charge excitations at the Fermi level in the system.
For a one-dimensional interacting electron gas, the low-energy charge
excitations are bosonic and have a velocity that is distinct from that of
the low-energy spin excitations.
The charge velocity $u_\rho$ depends on the bare
Fermi velocity and the interaction strength;
for $U=0$, both the spin and charge velocities reduce to the Fermi
velocity $v_F=2t\sin\left(\pi n\right)$.
For the one-dimensional Hubbard model, $u_\rho$ can be calculated for
all $n$ and $U$ using the Bethe ansatz \cite{schulz1993}.
These theoretical estimates as well as the light-cone velocities
obtained from fits to the disturbance in the local density in
Figs.~\ref{fig:u0n26v0}(a), \ref{fig:u4n26v0}(a), \ref{fig:u0n50v0},
and \ref{fig:u4n50v0}(a) are plotted 
in Fig.~\ref{fig:chargevel}.
Note that coupling the system to the probe reduces the local particle
density in the system.
We thus adjust the value of the average particle
density (abscissa of the plot) so that it corresponds to the average
value in the first trough within the light cone. 
We observe that all data points, except
the $U=0$ quarter-filled case, align well with the theoretical
estimates of the spreading speed of particles.

\clearpage

\end{document}


	%
	
	%
	\title{Time evolution of the local density of states of
          strongly correlated fermions coupled to a nanoprobe:
          supplemental material}
	\author{T. Blum}
	\affiliation{Fachbereich Physik, Philipps-Universit\"at Marburg, 
		35032 Marburg, Germany}
	\author{R.M. Noack}%
	\affiliation{Fachbereich Physik, Philipps-Universit\"at Marburg, 
		35032 Marburg, Germany}
	\author{S.R. Manmana}%
	\affiliation{Institut f\"ur Theoretische Physik, Georg-August-Universit\"at G\"ottingen, 37077 G\"ottingen, Germany}
	\maketitle
	
In this supplemental material, we make available the results of
additional calculations that elucidate some technical issues in our
numerical methods.
In particular, we address the following two issues: \emph{(i)} the issue of
convergence of the TDVP-based time-dependent DMRG calculations for the
interacting system and \emph{(ii)} the origin and form of oscillations
in the local density of states (LDOS) that are seen in the
noninteracting system.

\section{Convergence of the TDVP DMRG calculations for the interacting system}
        
\begin{figure}[!]
  \includegraphics[width=\columnwidth]{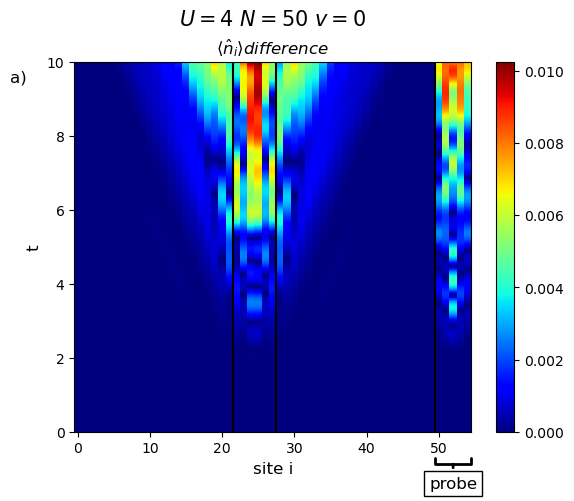}
  \includegraphics[width=\columnwidth]{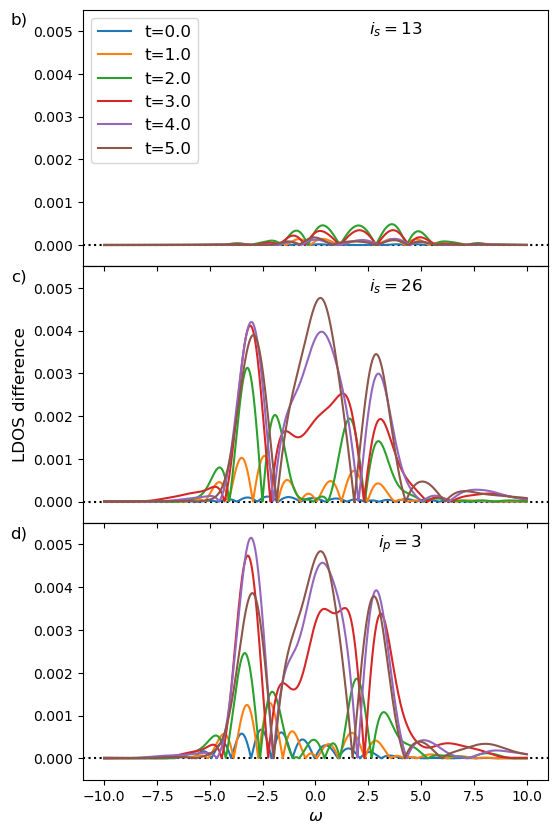}						
  \caption{Difference in results for the TDVP time evolution for
    maximum bond dimension $\chi=1500$ and $\chi=2500$ for a system with
    $L=50$ sample sites and $L_p=5$ nanoprobe 
    for $U=4$ and probe velocity $v=0$
    at half filling, i.e., $N=50$. 
    Here (a) depicts the absolute difference in the particle density
    $\langle n_i \rangle$, and (b)-(d) depict the absolute
    difference in the LDOS at the indicated sites.}	\label{fig:U4N50v0diff}				
\end{figure}

\begin{figure}[!]
  \includegraphics[width=\columnwidth]{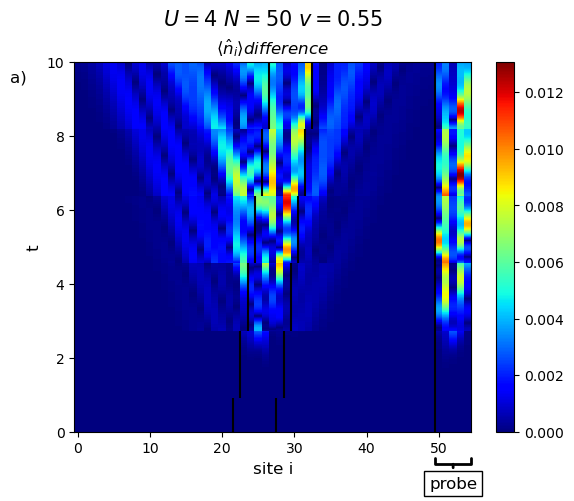}
  \includegraphics[width=\columnwidth]{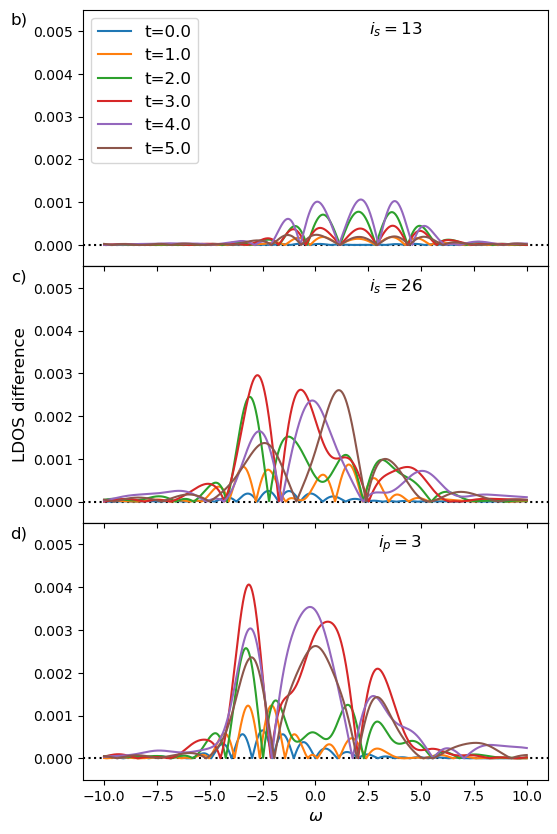}						
  \caption{Difference in results for the TDVP time evolution for
    maximum bond dimension $\chi=1500$ and $\chi=2500$ for a system with
    $L=50$ sample sites and $L_p=5$ nanoprobe 
    for $U=4$ and probe velocity $v=0.55$
    at half filling, i.e., $N=50$. 
    i.e., $N=50$. 
    Here (a) depicts the absolute difference in the particle density
    $\langle n_i \rangle$, and (b)-(d) depict the absolute
    difference in the LDOS at the indicated sites.}		\label{fig:U4N50v0.55diff}				
\end{figure}

Figs.~\ref{fig:U4N50v0diff} and Fig.~\ref{fig:U4N50v0.55diff} display the
difference in the local particle density \erwartung{n} and the LDOS of
the TDVP time evolution with a maximum bond dimension of $\chi=1500$
compared to the results shown in the paper, which were computed with a
maximum bond dimension of $\chi=2500$.
This can be regarded as a measure of the error of the MPS results.
In the $v=0$ case, Fig.~\ref{fig:U4N50v0diff}, the maximum
error in the local particle density, which occurs at later times (at
$t \gtrsim 7$)  is approximately $0.01$.
This error is approximately $2\%$ of the measured particle density for the
half-filled system.
The absolute error in the LDOS at site $i_s=12$, far away from the
probe, is well under $0.001$. 
The LDOS in the vicinity of the probe reaches an error of $0.005$ at
later times.
The peak of the LDOS, shown in the paper, has approximately the value
of $0.1$, so that the relative error at this point is approximately
about $5\%$.
However, note that this error takes only the dependence on the bond
dimension into account and not the limited resolution of the LDOS due
to the limited range of time in the integration.
Overall, we can say that the bond dimension of $\chi=2500$ seems to be
sufficient to compute results that are sufficiently accurate to
resolve major features in the LDOS such as the position and relative
weights of the stronger peaks reliably.

\begin{figure}[!]	
	\includegraphics[width=\columnwidth]{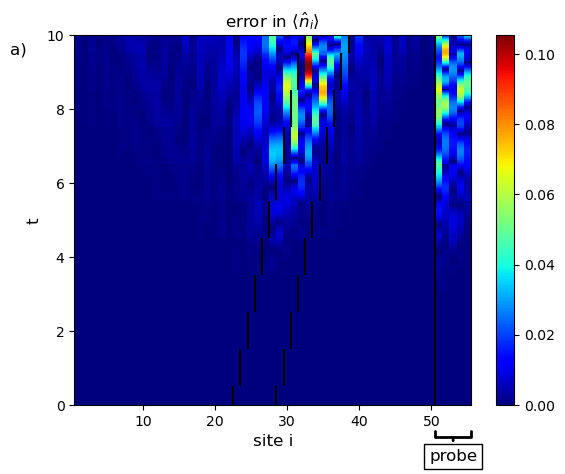}	
	\includegraphics[width=\columnwidth]{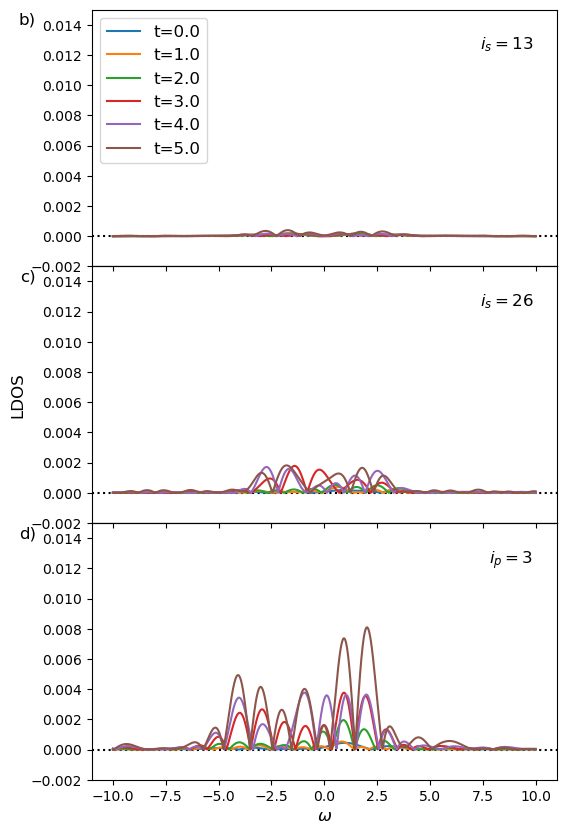}							
	\caption{
		Results of the TDVP time evolution, with $L=50$ sample sites and
		$L_p=5$ nanoprobe sites, compared to exact
		diagonalization for $U=0$ and probe velocity $v=1$ at half filling,
		i.e., $N=50$. 
		Here (a) depicts the absolute difference in the particle density
		$\langle n_i \rangle$, and (b)-(d) depict the absolute
		difference in the LDOS at the indicated sites.
		Note that the
		maximum integration time  %
		$\tau_{\text{max}}=5$ for both the
		exact diagonalization and TDVP calculations in order to make
		the results comparable.		
	} \label{fig:LDOS_error_v1}						
\end{figure}

To supplement to Fig. 12 in the main paper, which depicts the error in
the local particle density and in the LDOS in our simulation for the case of a 
stationary probe, $v=0$, for the noninteracting system, $U=0$, at half
filling, we include here a corresponding depiction 
of the error for a probe moving with velocity $v=1$,
also at $U=0$ and half-filling,
Fig.~\ref{fig:LDOS_error_v1}.
The error is comparable to or smaller than that in the
stationary case, equating to a maximum error of about
10\% for the largest peaks at later times.
Such accuracy should be sufficient
to resolve the qualitative structure of the LDOS for the moving
nanoprobe. 


\section{Oscillations in the LDOS}

\begin{figure}[!]
	\includegraphics[width=\columnwidth]{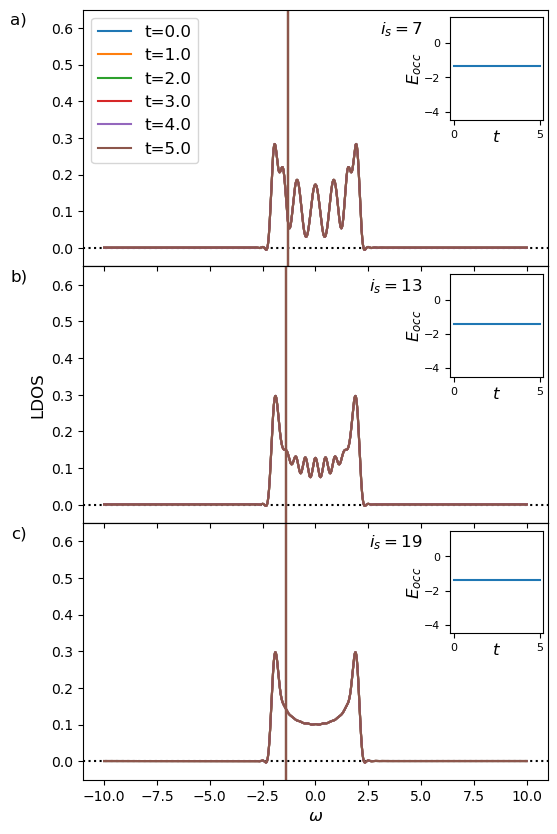}					
	\caption{
	Exact results for the Hubbard model
	with $L=50$ sample sites and no nanoprobe coupled to the Hubbard chain
	for the
	system initially at quarter filling, i.e., $N=26$, with
	$U=0$.
	Here
	(a), (b), and (c) display the LDOS %
	for the lattice sites $i_s=7$, $i_s=13$, and $i_s=19$,
	respectively,
	as a function of the frequency $\omega$.
	Note that the LDOS %
	is rigorously time-independent for $U=0$
	when $v=0$.
	The occupation energy
	$E_{\rm occ}(t)$ is shown as an inset %
	for each of the
	three lattice sites as a function of $t$ and is also
	shown at five indicated times
	as vertical lines in the LDOS.}	\label{fig:U0N26v0_no_coupling}					
\end{figure}

Fig.~\ref{fig:U0N26v0_no_coupling} displays the LDOS for an
isolated, noninteracting $L=50$ site chain, i.e., a chain with $U=0$
and no coupling to the nanoprobe for sites at three different
locations on the chain, $i_s=7$, a site close to the left end of the
chain, $i_s=13$, a site in the left half of the chain also treated in
the main paper for all systems, and a site near
the center of the system, at $i_s=19$.
At site $i_s=7$, the oscillations in the LDOS are much more pronounced than
at site $i_s=13$, and at $i_s=19$, they almost vanish altogether.
This is strong evidence that the cause of the oscillations in the LDOS
are due to end effects caused by the open boundaries of the system.

We now examine the behavior when the noninteracting $L=50$ chain is
coupled to the nanoprobe, first, for nanoprobe velocity $v=0$ and then
velocity $v=0.55$.
In particular, we want to examine the cause of an asymmetry in the
oscillations in the LDOS in the noninteracting system for the
stationary nanoprobe, as seen in Fig.~2(b) in the main text.

\begin{figure}[!]
	\includegraphics[width=\columnwidth]{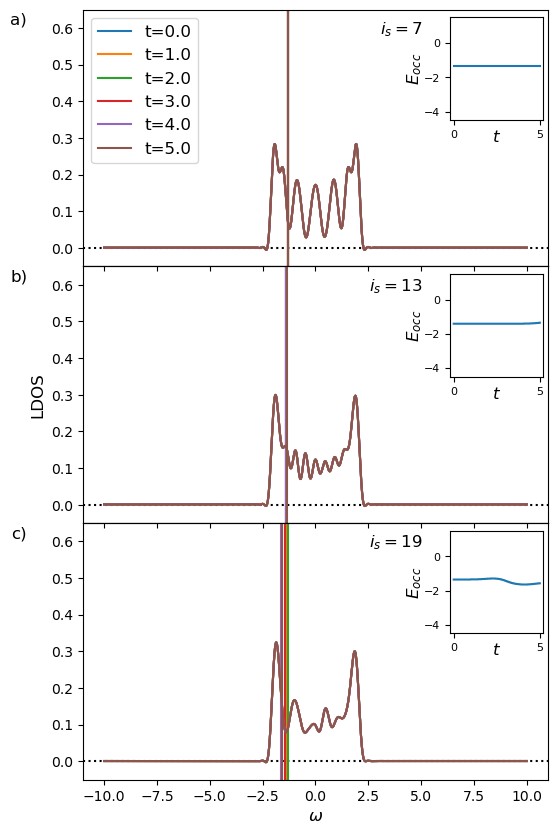}					
	\caption{Exact results for the nanoprobe model
		with $L=50$ sample sites and $L_p=5$ nanoprobe sites
		for the
		system initially at quarter filling, i.e., $N=26$, with
		$U=0$ %
		and nanoprobe %
		velocity $v=0$.
		Here
		(a), (b), and (c) display the LDOS %
		for the lattice sites $i_s=7$, $i_s=13$, and $i_s=19$,
		respectively,
		as a function of the frequency $\omega$.
		Note that the LDOS %
		is rigorously time-independent for $U=0$
		when $v=0$.
		The occupation energy
		$E_{\rm occ}(t)$ is shown as an inset %
		for each of the
		three lattice sites as a function of $t$ and is also
		shown at five indicated times
		as vertical lines in the LDOS.}		\label{fig:U0N26v0}				
\end{figure}

We note that the LDOS in Fig.~\ref{fig:U0N26v0_no_coupling} at site $i_s=13$ is
symmetric.
However, when the stationary probe is coupled to the system the LDOS at
site $i_s=13$ becomes asymmetric, as can be seen in
Fig.~\ref{fig:U0N26v0}(b).
This is a result of the interference of the effects of the edges of
both the sample and the nanoprobe.
At site $i_s=19$, Fig.~\ref{fig:U0N26v0}(c) this effect is much more
pronounced, whereas at site $i_s=7$ the coupling to the probe has no
visible effect.

\begin{figure}[!]
	\includegraphics[width=\columnwidth]{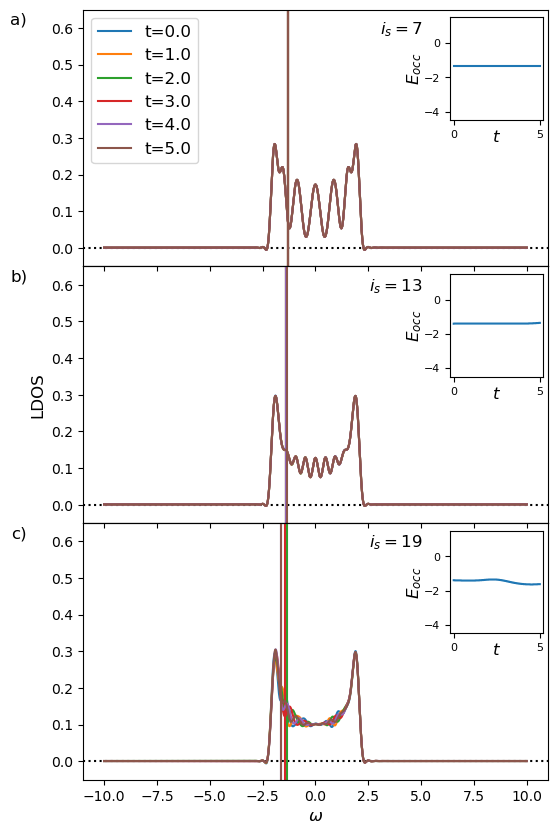}					
	\caption{Exact results for the nanoprobe model
		with $L=50$ sample sites and $L_p=5$ nanoprobe sites
		for the
		system initially at quarter filling, i.e., $N=26$, with
		$U=0$ %
		and nanoprobe %
		velocity $v=0.55$.
		Here
		(a), (b), and (c) display the LDOS %
		for the lattice sites $i_s=7$, $i_s=13$, and $i_s=19$,
		respectively,
		as a function of the frequency $\omega$.
		The occupation energy
		$E_{\rm occ}(t)$ is shown as an inset %
		for each of the
		three lattice sites as a function of $t$ and is also
		shown at five indicated times
		as vertical lines in the LDOS.                
                Note that the LDOS is rigorously time-independent
                for $U = 0$ \textit{only} when $v = 0$; thus, the
                time-dependence visible in the LDOS in (c) is not 
                not forbidden here.}	\label{fig:U0N26v0.55}
        
\end{figure}

For the moving nanoprobe, $v=0.55$, Fig.~\ref{fig:U0N26v0.55},
the LDOS at site $i_s=13$, Fig.~\ref{fig:U0N26v0.55}(b), remains
symmetric because the probe is moving away from this site, therefore
suppressing the effect the probe has on the LDOS within integration
time.
At site $i_s=19$,  Fig.~\ref{fig:U0N26v0.55}(b), we can see that the
LDOS is indeed affected by the coupling to the moving probe; however,
the effects are much smaller than in the stationary case.

\newpage
